\documentclass[fleqn,usenatbib]{mnras}
\usepackage{newtxtext,newtxmath}
\usepackage[T1]{fontenc}
\usepackage{ae,aecompl}
\usepackage{microtype}

\usepackage{graphicx}
\usepackage{amsmath}
\usepackage{amssymb}
\usepackage{xspace}
\usepackage{xcolor}
\usepackage[squaren]{SIunits}
\usepackage{scalefnt}
\usepackage{comment}

\hypersetup{colorlinks=true,allcolors=teal}

\def\bq{\begin{equation}} 
\def\eq{\end{equation}}
\newcommand{\bqa}{\begin{eqnarray}} 
\newcommand{\eqa}{\end{eqnarray}}

\defcitealias{Bird_2020}{B20}
\defcitealias{Hahn_2020}{H20}

\newcommand\smaller[2][0.85]{{\scalefont{#1}#2}}
\newcommand{\HACC}{\smaller{HACC}\xspace}
\newcommand{\HACCPM}{\smaller{HACC-PM}\xspace}
\newcommand{\HACCTPM}{\smaller{HACC-TPM}\xspace}

\newcommand{\MUSIC}{\textsc{monofonIC music-2}\xspace}
\newcommand{\SFSSOFTEN}{\smaller{SR-SOFTEN}\xspace}
\newcommand{\SFSOFF}{\smaller{SR-OFF}\xspace}
\newcommand{\SFSBOOST}{\smaller{SR-BOOST}\xspace}
\newcommand{\CAMB}{\smaller{CAMB}\xspace}

\newcommand{\NP}{N_p}
\renewcommand{\NG}{N_g}
\newcommand{\DeltaF}{\Delta_F}
\newcommand{\DeltaP}{\Delta_P}
\newcommand{\DeltaG}{\Delta_G}
\newcommand{\knyq}{k_{\rm Nyquist}}

\addunit{\massh}{\mathit{h}^{-1}M_\odot}
\addunit{\Mpch}{\mathit{h}^{-1}Mpc}
\addunit{\kpch}{\mathit{h}^{-1}kpc}
\addunit{\kms}{km\ s^{-1}}
\addunit{\invMpch}{\mathit{h}\ Mpc^{-1}}

\title[]{Numerical Discreteness Errors in Multi-Species Cosmological N-body Simulations}

\author[X. Liu et al.]{
\parbox[h]{0.95\textwidth}{
Xin Liu$^{1,2}$\thanks{E-mail: \href{mailto:xinliu8@uchicago.edu}{xinliu8@uchicago.edu}}, 
J.D. Emberson$^{3}$, 
Michael Buehlmann$^{1}$,
Nicholas Frontiere$^{3}$,
and Salman Habib$^{1,3}$}
\\
\\$^{1}$High Energy Physics Division, Argonne National Laboratory, Lemont, IL 60439, USA
\\$^{2}$Department of Physics, University of Chicago, Chicago, IL 60637, USA
\\$^{3}$Computational Science Division, Argonne National Laboratory, Lemont, IL 60439, USA
}

\date{Accepted XXX. Received YYY; in original form ZZZ}
\pubyear{2020}

\begin{document}
\label{firstpage}
\pagerange{\pageref{firstpage}--\pageref{lastpage}}
\maketitle

\begin{abstract}
We present a detailed analysis of numerical discreteness errors in two-species, gravity-only, cosmological simulations using the density power spectrum as a diagnostic probe. In a simple setup where both species are initialized with the same total matter transfer function, biased growth of power forms on small scales when the solver force resolution is finer than the mean interparticle separation. The artificial bias is more severe when individual density and velocity transfer functions are applied. In particular, significant large-scale offsets in power are measured between simulations with conventional offset grid initial conditions when compared against converged high-resolution results where the force resolution scale is matched to the interparticle separation.
These offsets persist even when the cosmology is chosen so that the two particle species have the same mass, indicating that the error is sourced from discreteness in the total matter field as opposed to unequal particle mass.
We further investigate two mitigation strategies to address discreteness errors: the frozen potential method and softened interspecies short-range forces. The former evolves particles under the approximately ``frozen'' total matter potential in linear theory at early times, while the latter filters cross-species gravitational interactions on small scales in low density regions. By modeling closer to the continuum limit, both mitigation strategies demonstrate considerable reductions in large-scale power spectrum offsets.  
\end{abstract}

\begin{keywords}
cosmology: theory --- large-scale structure of Universe --- methods: numerical --- gravitation
\end{keywords}


\section{Introduction}
\label{sec:introduction}

Numerical simulations of cosmological structure formation are an indispensable tool for providing theoretical predictions of the distribution of matter on non-linear scales. In combination with observational measurements, these predictions can be used to provide an understanding of structure formation and to solve the inverse problem of determining the cosmological makeup of the Universe. Next-generation observational 
probes -- including DESI \citep{desi:2016}, the Roman Space Telescope~\citep{roman:2015}, Rubin Observatory's LSST \citep{lsst:2019}, Euclid \citep{euclid:2011}, and SPHERE$^{\rm{x} }$~\citep{spherex:2014} -- will continue to refine measurement constraints and thus require a matched increase in precision from theoretical modeling. From this point of view, there are two primary avenues that need to be advanced in simulations: 1) increased physical fidelity, including the accurate treatment of hydrodynamic forces and astrophysical processes impacting baryons, as well as inclusion of ingredients such as massive neutrinos, early dark energy, modified gravity, and primordial non-Gaussianity; 2) an improved understanding of the systematic errors inherent to the simulation approach, particularly with regard to the multi-species case that is both more numerically complex and less  studied. 

The earliest and most common form of cosmological simulation uses the N-body approach {\citep[e.g.,][]{klypin1983,efstathiou/etal:1985,barnes/hut:1986,1988book}},
which can be viewed as a dynamical Monte Carlo algorithm applied to the Vlasov-Poisson system of equations in an expanding universe, for which a set of point particles sampling the total matter field are evolved under their mutual gravitational potential. Each particle represents a parcel of the total matter fluid with a macroscopic mass that scales as the cube of the mean interparticle separation, $m_p \propto \DeltaP^3 \equiv (L/\NP)^3$, where $L$ is the simulation box width and $\NP$ is the number of particles along one dimension. It is common to solve the equations of motion using a gravitational force resolution, $\DeltaF$, that is finer than the mass resolution scale set by the mean interparticle separation, $\DeltaP$. In this case, it is natural to question whether the inherent discreteness of the particle representation leads to a numerical solution that systematically departs from the desired continuum limit. In the language familiar from plasma physics, we are interested in the circumstances under which the Klimontovich description (corresponding to the fine-grained distribution function) agrees with the mean field description underlying the Vlasov equation~\citep{nicholson}. We note that our concern here is not so much with particle collisionality effects~\citep{okuda1970} which may be of concern in the deeply non-linear region, but with issues arising at much larger scales.

This question has received significant attention with various studies addressing the problem from a number of perspectives \citep[e.g.,][]{melott/etal:1997,splinter/etal:1998,knebe/etal:2000,power/etal:2003,diemand/etal:2004,heitmann2005ApJS, romeo/etal:2008,joyce/etal:2009, heitmann2010ApJ}. The viewpoint normally taken is that discreteness errors are indeed introduced when $\DeltaF \lesssim \DeltaP$, but that these 1) remain confined to small scales due to the preferential transfer of power from large to small scales in the cold dark matter (CDM) paradigm, and 2) are effectively washed out at later times due to non-linear growth and mixing \citep{little/etal:1991,hamana/etal:2002,bagla/prasad:2009}. Note that this is not necessarily the case in warm dark matter (WDM) or hot dark matter (HDM) scenarios where the lack of small-scale cosmological power is unable to mask discreteness errors, leading to obvious numerical artifacts that survive to late times \citep{gotz/sommerlarsen:2003,wang/white:2007,melott:2007,power/etal:2016}. 

Discreteness errors become more readily apparent in particle-based simulations with multiple species as the symmetry amongst particles is broken and additional metrics can be analyzed. For instance, if the standard N-body method is modified so that the total matter field is split into two species of differing masses, discreteness errors manifest in the form of mass segregation within collapsed objects as energy is transferred from the heavy to the light species \citep{efstathiou/eastwood:1981,binney/knebe:2002,JD2019}. Such two-body interactions further extend to an artificial heating of gas \citep{steinmetz/white:1997} in the situation where hydrodynamic forces are additionally modeled. In this case, one must also consider the more general interplay between the gravitational and hydrodynamic force resolutions to avoid numerical convergence issues, particularly when sub-resolution physics treatments are used \citep{bate/burkert:1997,hopkins/etal:2018,ludlow/etal:2019}. In addition to modeling CDM and baryons, further discreteness effects occur when simulating neutrino particles. Recently, \citet{sullivan/etal:2023} showed that the discrete signature of the initial condition neutrino particle grid can transfer onto the total matter field if the neutrino resolution is too coarse. Fortunately, in each of the above cases, the discreteness errors are mostly confined to small scales, making them easier to isolate and control.

Cosmological hydrodynamic simulations \citep[e.g.,][]{illustris:2014,eagle:2015,bahamas:2017,JD2019} tend to make the approximation that both the CDM and baryon tracer particles sample the total matter field (i.e., they are both initialized using the mass-weighted sum of the individual CDM and baryon transfer functions). This is technically incorrect since the disparate early universe evolution of the two species leads to markedly different transfer functions at the common initial redshifts ($z \approx 50-200$) of simulations. However, \citet{angulo2013} showed that using the physically consistent procedure of initializing each species with its respective transfer function introduces a new discreteness error that, unlike the cases above, propagates to large scales. Specifically, if the particles are evolved with a force resolution $\DeltaF \lesssim \DeltaP$, then both the CDM and baryons acquire incorrect growth that extends all the way out to the largest scales at which the two transfer functions are initially different ($k \gtrsim \unit{10^{-2}}{\invMpch}$). The incorrect growth operates in opposite directions for the two species and has a greater amplitude in the lighter baryon particles. \citet{angulo2013} attribute this problem to spurious particle coupling between the unequal mass species and suggest using an adaptive softening length in the CDM-baryon gravitational force interactions as a solution. Similar conclusions were also reached by \citet{oleary/mcquinn:2012}.

Alternative solutions to this problem have been proposed by \citealt{Bird_2020} and \citealt{Hahn_2020}, hereafter referred to as \citetalias{Bird_2020} and \citetalias{Hahn_2020}, respectively. The preferred method in \citetalias{Bird_2020} is to switch from the conventional offset grid pre-initial condition (pre-IC) setup to a mixed grid$+$glass \citep{glass_note} configuration. Another stated solution is to maintain an offset grid configuration, but to under-sample the baryon particles relative to the CDM so that each species has the same mass. In other words, the findings in \citetalias{Bird_2020} are consistent with the idea that the discreteness error arises from the fact that the particle masses are unequal. The method presented in \citetalias{Hahn_2020} similarly involves modifications to the ICs. In this case, an offset grid configuration can still be used, but the individual transfer functions are encoded in mass perturbations instead of positional displacements. While the proposed methods in \citetalias{Bird_2020} and \citetalias{Hahn_2020} successfully diminish the spurious growth of power, a complete theoretical explanation of both the propagation of discreteness errors and how the respective solutions formally mitigate these effects in a controlled way remains to be worked out.

In this paper, we continue the investigation into the origin of the large-scale discreteness error in mixed CDM plus baryon simulations. This is achieved using a controlled set of multi-species simulations that focus exclusively on gravitational forces so as to avoid the extra symmetry breaking that occurs when baryons are subject to hydrodynamic forces. We isolate for the impacts of having unequal masses versus unequal transfer functions by running simulations where either the mass or transfer function is held common for both species while the other is varied. We find that the discreteness error still exists in the case where individual transfer functions are used but the cosmology is chosen so that the particle masses are the same. In this case, the only difference compared to the standard cosmology is that the amplitude of the incorrect growth is shared equally between the CDM and baryon particles rather than mainly in the baryons. We argue that the discreteness error is not driven by having unequal particle masses, but more generally due to an unmixed discretization of the total matter field. In addition, this error is fundamentally distinct from artificial particle coupling that leads to small-scale features such as mass segregation.

In an effort to improve the discreteness error in multi-species cosmological simulations, we also evaluate various mitigation strategies. In addition to the proposed strategies in \citetalias{Bird_2020} and \citetalias{Hahn_2020}, as well as starting the simulation at later times, we propose two new strategies: 1) an approach based on evolving the particles under the linear total matter field at early times, and 2) using softened cross-species force interactions in low-density regions. 
We show that the biased power growth can be significantly reduced with either mitigation strategy.

The rest of the paper is organized as follows. We present an overview of our numerical methods and relevant background information in Section~\ref{sec:method}. In Section~\ref{sec:unequalmass}, we focus on simulations that use the standard approach of initializing the CDM and baryon components with the total matter transfer function and show that the discreteness error exists in this case, but is confined only to small scales. We then switch to the physically consistent initialization strategy in Section~\ref{sec:unequaltransfer} where the two species are initialized with their respective transfer functions. Here we confirm previous findings that this leads to large-scale biases in the growth of both species and demonstrate that this occurs even if the particles have the same mass. This is further illustrated in Appendix~\ref{app:planewave} using a simple two-fluid plane wave toy model.
In Section~\ref{sec:convergence}, we compare the proposed solutions from \citetalias{Bird_2020} and \citetalias{Hahn_2020} and also present two new mitigation strategies to suppress the discretization error. Concluding remarks and further discussions are presented in Section~\ref{sec:conclusion}.

\section{Method}
\label{sec:method}

We provide below a summary of the numerical methods used in this work. We begin in Section~\ref{subsec:sims} with a description of the cosmological simulations performed here, followed in Section~\ref{subsec:ics} with further details on the multi-species ICs used to initialize each run. We finish in Section~\ref{subsec:fp} with an outline of the frozen potential method that is later presented as a potential mitigation strategy for large-scale discreteness errors.

\subsection{Simulations}
\label{subsec:sims}

The simulations in our paper are performed using the Hardware/Hybrid Accelerated Cosmology Code (\HACC). \HACC is a cosmological N-body solver designed to scale efficiently on all modern supercomputing platforms. We will briefly describe below the main code features of \HACC relevant to this work and refer the reader to \citet{Habib2016} for more specific details. Despite the fact that we carry out mixed CDM plus baryon simulations, we restrict attention to purely gravitational forces so that we can easily identify particle asymmetries arising from discretization. In addition, we consider two different types of gravitational force solvers that are included in \HACC to illustrate that discreteness errors arise regardless of the nature of the local force algorithm. 

In the first method, which we refer to as \HACCPM, the gravitational force is computed using a spectral solver that employs a particle mesh (PM) algorithm to solve the Poisson equation in Fourier space. In this approach, particle masses are deposited to a uniform mesh containing $\NG^3$ cells via a cloud-in-cell (CIC) interpolation. In the second method, which we refer to as \HACCTPM, the PM code is used to evaluate a long-range gravitational force and is further augmented with a short-range force solver that traverses a tree data structure \citep{gafton&rosswog:2011} to compute direct pairwise gravitational interactions on small scales. In this case, a spectrally filtered ``quiet'' PM algorithm is matched to the short-range force at a handover scale, $r_h$, defined to equal roughly three grid separations:
\begin{equation}
\label{equ:hand_over_scale}
r_h \approx 3\frac{L}{\NG} \equiv 3 \DeltaG.
\end{equation}
The short-range force is evaluated only for particle pairs confined to the handover scale and, for sufficiently distant leaves (determined via an opening angle criterion), a superparticle at the leaf center of mass is used instead of looping over all particles on the leaf.
The short-range force is smoothed on small scales using a tunable Plummer softening to avoid artificially strong particle interactions. To expedite the particle search, we build a chaining mesh \citep[CM;][]{1988book} containing cells of width $l_{\rm CM} = 4\DeltaG$, chosen to encapsulate $r_h$, and construct an independent tree (``bushes'') within each cell.

We will consider simulations with a range of force resolutions in order to show that discreteness errors arise when the force resolution is made finer than the particle separation. In each run, we simulate $2\times\NP^3$ CDM plus baryon particles and define the mean interparticle separation of each species as $\DeltaP = L/\NP$. In the case of \HACCPM, the force resolution is set by the PM grid spacing, $\DeltaG$, which we adjust by varying $\NG$. For the \HACCTPM runs, we fix the PM mesh so that $\NG = \NP$ and instead vary the force resolution by setting the softening length, $r_{\rm soft}$, to be an adjustable fraction of the grid scale. In other words, the force resolutions of the two simulation methods can be expressed in terms of the mean interparticle separation as:
\begin{equation}
\DeltaF = \epsilon \DeltaP,
\end{equation}
where $\epsilon = \NP/\NG$ for \HACCPM and $\epsilon \approx r_{\rm soft}$ for \HACCTPM. We will show that discreteness errors arise whenever $\epsilon \lesssim 1$.

Throughout the paper, we assume a $\rm \Lambda CDM$ model with the following cosmological parameters:
\begin{equation}
\begin{aligned}
\Omega_m&=\Omega_c+\Omega_b=0.3, \ \Omega_{\Lambda}=0.7,\\ h&=0.7, \ \sigma_8=0.8, \ n_s=0.96.
\end{aligned}
\end{equation}
The corresponding mass of each simulation particle, $m_{\alpha}$, is set by the mean density of the Universe, $\bar{\rho}$, mean interparticle separation, $\DeltaP$, and energy density, $\Omega_{\alpha}$, of the given species:
\begin{equation}
m_{\alpha} = \bar{\rho} \Omega_{\alpha} \DeltaP^3,
\end{equation}
with $\alpha \in \{c, b\}$ denoting the CDM or baryon component, respectively. We consider two scenarios in our tests below: 1) an equal density cosmology with $\Omega_c=\Omega_b=0.15$ (yielding a particle mass ratio, $q \equiv m_c/m_b = 1$, of the simulated CDM and baryon particles), and 2) $\Omega_c=0.25$ and $\Omega_b=0.05$, referred to as the ``concordance cosmology'' ($q=5$). The equal density case is used in Section~\ref{subsec:uniformcosmo} to investigate discretization errors in the absence of particle mass disparities. The concordance cosmology is used in all remaining runs and was designed to approximately match the energy density ratio observed with Planck \citep{Planck2020}. 

\subsection{Initial Conditions}
\label{subsec:ics}

The general strategy for creating cosmological ICs is to start with a random realization of the target density field. This is achieved in Fourier space by taking the point-wise product of a Gaussian random white noise field, $\mu(\textbf{k})$, and the square root of the (dimensional) power spectrum:
\begin{equation}
\label{equ:delta}
\delta_{\alpha}(\textbf{k},z_i)=\mu(\textbf{k})\sqrt{P_{\alpha}(k,z_i)},
\end{equation}
with $\alpha \in \{m, c, b\}$, where $m$ denotes the total matter field.
Here $z_i$ is the initial redshift of the simulation which is taken to be $z_i=200$ for the majority of runs considered here.
The power spectrum can be written in terms of the linear theory (LT) transfer function, $T_\alpha$, in the following manner:
\begin{equation}
P_{\alpha}(k,z_i)=Ak^{n_s}T_{\alpha}^2(k,z_i),
\label{eq:Palpha}
\end{equation}
where $n_s$ is the spectral index and $A$ (not to be confused with the scalar amplitude $A_s$) is a normalization constant set by the value of $\sigma_8$.

There are generally two conventions in multi-species simulations for the choice of the target density field. The most common case is to use the total matter transfer function to initialize both the CDM and baryon particles with an implicit assumption that baryons follow the dark matter. We refer to this as the ``single-fluid'' case since both the CDM and baryon particles act as tracers of the same total matter fluid despite the fact that the two species are meant to individually represent the separate components of that field. The more formally consistent approach, especially at high redshifts, is to use the individual CDM and baryon transfer functions to initialize each respective species; we refer to this as the ``two-fluid'' case.

Transfer functions are normally computed using an Einstein-Boltzmann code \citep[e.g., CAMB;][]{Lewis2000}. However, one needs to be careful of differences in the forward modeling assumed in cosmological simulations compared to that of Boltzmann codes. In particular, simulations implicitly work in the Newtonian framework of gravity, whereas Einstein-Boltzmann codes also include other physical effects such as the coupling between matter and radiation. Accordingly, the usual approach is to ``back-scale'' the transfer function evaluated at the final redshift of the simulation, $z_f$, to the starting redshift using the same forward modeling used in the simulation. In the single-fluid case,
this process involves simple multiplicative factors: 
\begin{equation}
\label{equ:transfer_m}
T_m^{\rm bs}(k,z_i)=\frac{D(z_i)}{D(z_f)}T_m(k,z_f),
\end{equation}
where $D(z)$ is the linear growth factor of the simulation forward model, normalized so that $D(0)=1$, and the superscript ``bs'' denotes the back-scaled result. This back-scaling strategy follows trivially from the fact that, in linear theory, the total matter density field evolves in a scale-independent manner, $\delta_{m}(\textbf{k},z)\propto D(z)$, in the Newtonian forward model.

This procedure needs to be modified in the two-fluid case since the density perturbations of CDM and baryons are both scale-dependent, even in the Newtonian forward model. For instance, baryons are coupled to photons via Compton scattering before recombination which suppresses the growth of the perturbations compared to CDM and also imprints baryon acoustic oscillations (BAO) on the baryon field. After recombination, the scale-dependent differences in the CDM and baryon density fields gradually transfer onto each other, though significant disparities still remain at the common starting redshifts used in simulations. The strategy used here is to take the logical extension of the standard back-scaling procedure in Eq.~(\ref{equ:transfer_m}) one step further:
\begin{equation}
\label{equ:transfer_ind}
T_{\alpha}^{\rm bs}(k,z_i)=\frac{T_{\alpha}(k,z_i)}{T_m(k,z_i)}T_m^{\rm bs}(k,z_i),
\end{equation}
where $\alpha=\{c,b\}$. In this way, we maintain the same total matter field as in the single-fluid case while ensuring that the relative proportions of the CDM and baryon density fields match the Einstein-Boltzmann predictions at $z_i$.

We use the back-scaled transfer functions, $T_\alpha^{\rm bs}$, to construct a random realization of the target density field following Eq.~(\ref{equ:delta}). This is evaluated on a uniform mesh containing $\NP$ cells along each axis. Next, we use the Zel'dovich approximation \citep[ZA;][]{Zeldovich1970} to compute a displacement field, $\textbf{d}$, that maps particles from their Lagrangian location, $\textbf{q}$, to their starting position, $\textbf{x}$, in the simulation:
\begin{equation}
\label{eq:lagran_q}
\textbf{x}(\textbf{q},z_i)=\textbf{q}+\textbf{d}(\textbf{q},z_i).
\end{equation}
The displacement field is derived from the density field in Fourier space \citep[e.g.,][]{white:2014}:
\begin{equation}
\textbf{d}_{\alpha}(\textbf{k},z_i)=\frac{i\textbf{k}}{\lVert\textbf{k}\rVert^2}\delta_{\alpha}(\textbf{k},z_i).
\label{eq:disp_final}
\end{equation}
Unless otherwise specified, we use a pre-IC configuration where the CDM and baryon particles are placed on uniform grids that are offset from each other by $\DeltaP/2$. In this way, the two particle grids are maximally offset from each other; this configuration is chosen to minimize any artificial particle coupling at early times \citep{Yoshida_2003}. In addition, we offset each particle grid by an amount $\DeltaP/4$ from the vertices of the IC mesh used to sample the displacement field. This is done to ensure that the two species are initialized in a symmetric manner with respect to the IC mesh (as well as the PM mesh used later in the force solver) so that the particle-grid relationships do not contribute to any of the particle asymmetries analyzed later. The displacement field is then spectrally interpolated at the particle positions \citep[see][]{valkenburg/etal:2017}. For glass configurations, we instead use an inverse-CIC interpolation in real space. In this case, we additionally deconvolve the displacement field with the CIC filter in order to compensate for interpolation smoothing.

The initial particle velocity is computed from the time derivative of Eq.~(\ref{eq:lagran_q}) which, along with Eq.~(\ref{eq:disp_final}), yields
\begin{equation}
\label{eq:vel_all}
\textbf{v}_{\alpha}(\textbf{k},z_i)=\frac{i\textbf{k}}{\lVert\textbf{k}\rVert^2}\left.\frac{\partial \delta_{\alpha}(\textbf{k},z_i)}{\partial t}\right|_{t=z_i}.
\end{equation}
In the single-fluid case, we can use the scaling $\delta_{m}(\textbf{k},z)\propto D(z)$ to show that 
\begin{equation}
\textbf{v}_{\alpha}(\textbf{k},z_i)= \frac{\dot{D}(z_i)}{D(z_i)}\textbf{d}_{\alpha}(\textbf{k},z_i).
\end{equation}
Therefore, the velocity is in the same direction as the displacement field with a magnitude that scales with the time derivative of $D(z_i)$. In other words, the initialization strategy of standard single-fluid simulations requires knowledge of only $T_m$, $D$, and $\dot{D}$; the latter two of which can be numerically evaluated for any given cosmology.

On the other hand, the two-fluid approach requires that we explicitly compute the scale-dependent time derivative of $\delta_\alpha$ in Eq.~(\ref{eq:vel_all}). In this case, it is convenient to define a velocity transfer function, $T_{\theta,\alpha}(k,z)$, where $\theta \equiv \nabla \cdot \textbf{v}$ denotes the velocity divergence field. The velocity transfer function follows from the linearized continuity equation \citep[e.g.,][]{peebles:1993} as the negative of the time derivative of the density transfer function:
\begin{equation}
T_{\theta,\alpha}(k,z) \equiv -\left.\frac{dT_{\alpha}(k,z)}{dt}\right|_{t=z}.
\end{equation}
In practice, the velocity transfer function can be computed using a finite difference of the synchronous gauge density transfer function sampled at $z \pm \Delta z$ \citep{valkenburg/etal:2017}. For consistency, this must be performed using the back-scaled density transfer functions:
\begin{equation}
T_{\theta,\alpha}^{\rm bs}(k,z_i) = \frac{H(z_i)}{a(z_i)}\frac{T_{\alpha}^{\rm bs}(k,z_{i}+\Delta z)-T_{\alpha}^{\rm bs}(k,z_{i}-\Delta z)}{2\Delta z},
\label{eq:tbstheta}
\end{equation}
where we choose $\Delta z = 1.0$. The velocity transfer function is used to generate a random realization of the velocity field, $\theta_\alpha(\textbf{k},z_i)$,  following Eqs.~(\ref{equ:delta}) and (\ref{eq:Palpha}), with the same white noise field and normalization constant that were used for the density field. The initial particle velocities become
\begin{equation}
\textbf{v}_{\alpha}(\textbf{k},z_i)=-\frac{i\textbf{k}}{\lVert\textbf{k}\rVert^2} \theta_\alpha(\textbf{k},z_i),
\end{equation}
where the negative sign arises from the definition of the velocity transfer function.

\begin{figure}
    \centering
    \includegraphics[width=\linewidth]{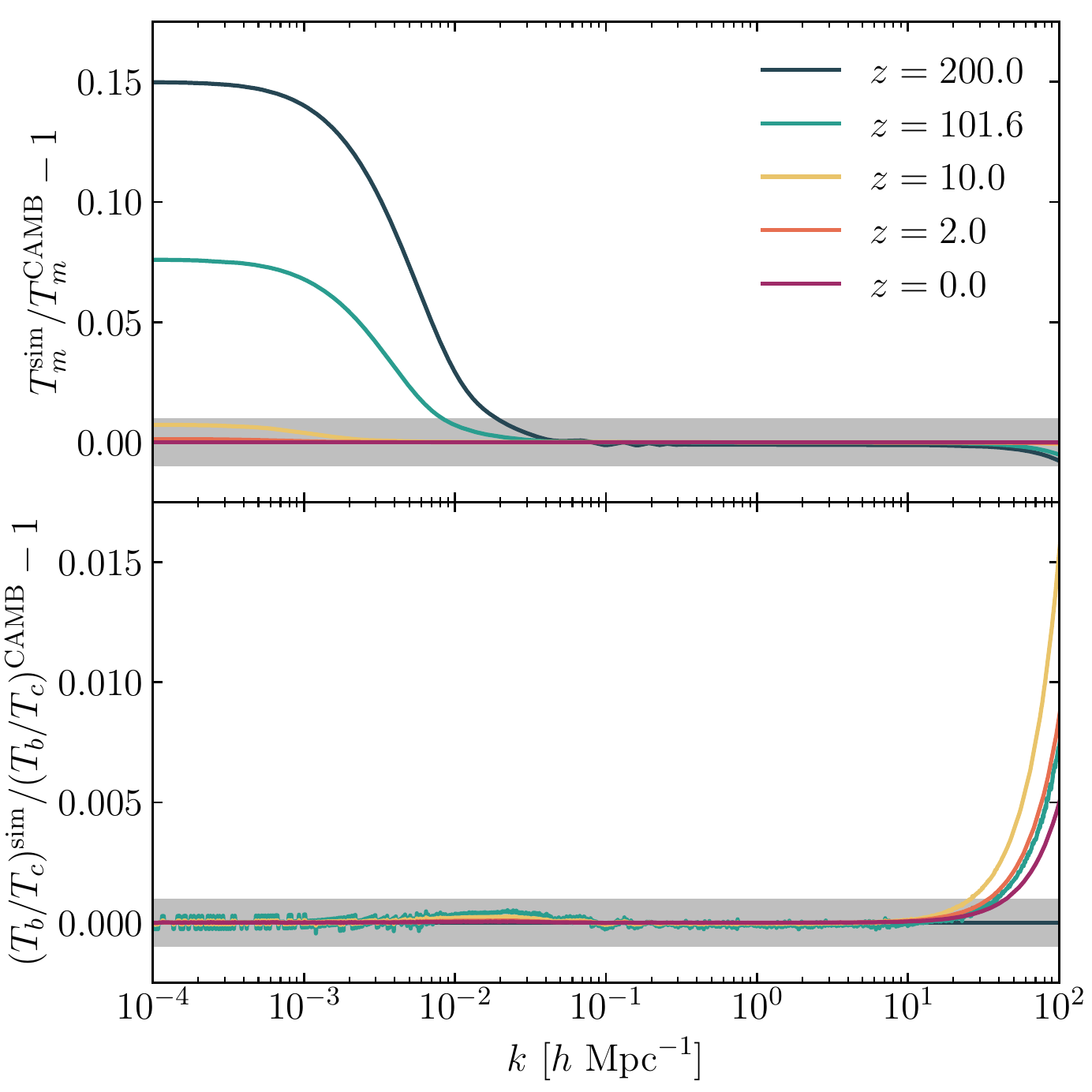}
    \caption{Top panel compares the linear total matter transfer function obtained in the simulation to \CAMB at various redshifts between $z_i = 200$ and $z_f = 0$. Similarly, the bottom panel compares the ratio of the linear baryon-to-CDM transfer functions from the simulation and \CAMB. Shaded regions denote $1\%$ and $0.1\%$ agreement in the top and bottom panels, respectively. The small amount of noise seen at $z = 101.6$ in the bottom panel arises from discreteness in the finite difference used to compute the initial velocity transfer function.}
    \label{fig:ics}
\end{figure}

We check for self-consistency in our two-fluid back-scaling approach by numerically integrating the initial CDM plus baryon density and velocity transfer functions defined in Eqs.~(\ref{equ:transfer_ind}) and (\ref{eq:tbstheta}) from $z_i = 200$ to $z_f = 0$ using the linearized Euler, continuity, and Poisson equations. The results are displayed in Fig.~\ref{fig:ics}. In the top panel, we show the relative difference between the total matter transfer function inferred by \HACC and that computed by \CAMB at a variety of redshifts between $z_i$ and $z_f$. At early times, the simulated total matter density field will have systematically higher power on large sales. This difference gradually diminishes with decreasing redshift until the total matter transfer function matches \CAMB on all scales at $z_f = 0$. This is the intended result of the back-scaling procedure and matches the usual behavior exhibited in standard single-fluid simulations. In the bottom panel, we compare the ratio $T_b/T_c$ of the simulation to that predicted by \CAMB. By construction, this ratio matches \CAMB at $z_i$ and maintains excellent agreement at the $0.1\%$ level at all times for scales $k \lesssim \unit{30}{\invMpch}$. On smaller scales, we do observe a systematic departure at lower redshift which reflects the omission of Jeans damping in the simulation forward model. In any event, this discrepancy is rather small (at the $1\%$ level) and confined to scales that are heavily impacted by non-linearities and baryonic physics at late times.

\subsection{Frozen Potential}
\label{subsec:fp}

We show later that discreteness errors manifest on large scales for two-fluid simulations run at high force resolution. The amplitude of these errors is greatest at high redshift when weak particle mixing amplifies the discrete nature of the simulated density field. An obvious way to reduce these errors is to start the simulation at lower redshift, but this is more difficult in the two-fluid case since no fully self-consistent higher-order strategy beyond the ZA currently exists (see \citetalias{Hahn_2020} for a method that is second- or third-order in displacement and first-order in relative velocity). We will present an alternative approach based on the frozen potential (FP) approximation.

The FP method as an approximation to the full non-linear problem was introduced by \citet{brainerd:1993} and \citet{Bagla_1994} and is named after the fact that the linear gravitational potential, $\phi \propto \delta_m/a$, remains constant in an Einstein-de~Sitter (EdS) universe where $\delta_m \propto a$. (For a further discussion and comparison with other approximate methods, see~\citealt{vrastil:2020}.) We use a modification to this scheme that accounts for the fact that the potential is not frozen in a general $\Lambda$CDM universe, but rather scales as $D/a$. The idea is that we first construct ICs in the usual manner described above with the additional step of storing in memory the initial gravitational potential,
\begin{equation}
\phi(k, z_i) = -\frac{4\pi G \bar{\rho}}{k^2} \delta_m(k, z_i),
\end{equation}
expressed here in Fourier space with comoving units. In this expression, $\delta_m(k, z_i)$ is the total matter field of the given realization; i.e., evaluated with Eq.~(\ref{equ:delta}). Then, instead of evolving
the particles under self-gravity, we scale $\phi(k,z_i)$ by $D/a$ at each time step and use this to compute the gravitational acceleration, $\textbf{g} = -\nabla \phi$, evaluated on the PM mesh. In other words, we replace the discrete gravitational potential computed using the simulation particles with the continuous gravitational potential predicted by the linear total matter field. For consistency, short-range particle forces are omitted. This ``frozen'' potential method is used until some later handover redshift, $z_h$, chosen large enough that LT still mainly holds. Afterwards, the particles are evolved until $z_f$ using the usual force solver based on the particle potential field.

\begin{figure}
    \centering
    \includegraphics[width=\columnwidth]{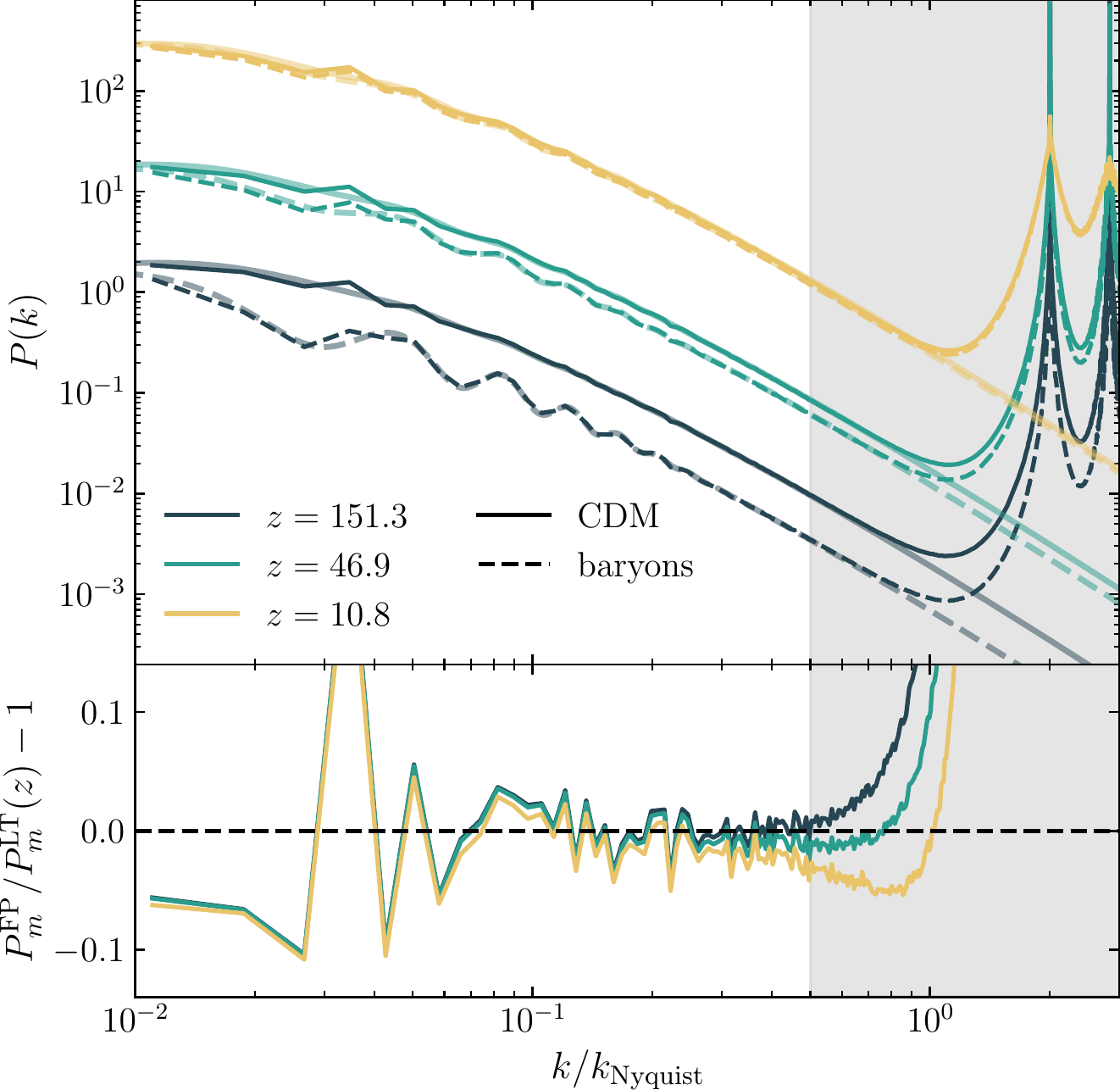}
    \caption{Power spectra from a FP simulation performed with $2\times256^3$ particles in a $\unit{512}{\Mpch}$ box compared to LT predictions. The top panel shows CDM (solid lines) and baryon (dashed lines) power spectra from the FP run (thick lines) and LT (shaded lines) at different redshifts denoted in the legend. The bottom panel shows the total matter power spectrum ratio between FP and LT at different redshifts. The vertical shaded gray band in each panel indicates scales above half the particle Nyquist frequency.}
    \label{fig:frozen_potential}
\end{figure}

In Fig.~\ref{fig:frozen_potential} we compare LT power spectra to the output of a FP simulation evolved to $z_h = 10.8$. The FP run is performed in a $\unit{512}{\Mpch}$ box containing $2\times256^3$ CDM plus baryon particles with an $\NG = 1024$ PM grid. As can be seen in the top panel, the FP simulation accurately reproduces the LT power spectrum for both CDM and baryons on all scales down to roughly half the particle Nyquist frequency, $k \lesssim \knyq/2$, where $\knyq \equiv \pi/\DeltaP$. On smaller scales, the simulated power spectra systematically deviates from LT, owing to the finite particle resolution of the simulation. Prominent spikes are also seen on small scales, with the first at $2\knyq$, which simply trace the discrete Fourier modes of the IC particle grid. The total matter power spectrum ratio between the FP run and LT is shown in the bottom panel. All three curves overlap on large scales with fluctuations around the LT prediction. These fluctuations are imprinted from the random white noise field used to seed the ICs and denote the sample variance associated with the finite simulation volume. While the white noise profile of the total matter field remains relatively fixed during evolution (as demonstrated by the overlap of each curve in the lower panel), this is not quite the case for the individual CDM and baryon power spectra due to the scale-dependent mixing of the two species.

In Section~\ref{subsec:mitfp}, we treat the FP method as an alternative to using higher-order Lagrangian Perturbation Theory (LPT) to initialize two-fluid simulations at lower redshift. In this way, we start with the ZA (i.e., first order LPT) at $z_i$ and then evolve the particles under the FP method until $z_h$. We thus effectively extend the IC from $z_i$ to $z_h$, but in an improved manner over simply applying the ZA at $z_h$. The improvement comes from the fact that particle motion responds to the local potential at each timestep during the FP evolution, enabling particle trajectories to acquire curvature as opposed to remaining on straight lines with the ZA. As a result, the FP method better resolves shell crossing at later times and provides a more accurate treatment of the linear growth of small-scale structure \citep[see][]{vrastil:2020}.
The added utility of the FP method for the purposes discussed here is that it uses a continuous representation of the total matter field and therefore circumvents errors arising from the discrete particle field. The FP method is also computationally efficient since it only needs to evaluate the real-space gravitational force field, $\textbf{g}(z_i)$, once and scales this result with $D/a$ at subsequent time steps. Of course, the FP approximation is based entirely on LT meaning that care must be taken on the choice of $z_h$. In addition, the force resolution is limited on small scales by the size of the PM mesh.

\section{Single-fluid Simulations}
\label{sec:unequalmass}

In this section, we present the results of simulations where two species of unequal mass are initialized with the same total matter transfer function. This is the common approximation made in cosmological hydrodynamic simulations where separate CDM and baryon species are evolved from an initial realization of their combined field. 

We begin in Section~\ref{subsec:sferror} with a demonstration that this setup leads to small-scale discreteness errors that manifest early during the simulation. The amplitude of this error increases with the mass ratio of the particles, but remains bounded in the asymptotic limit when the mass ratio approaches infinity (i.e., when one species becomes a passive tracer of the other species). We argue that this error arises from a coherent discretization of the total matter field into an unmixed representation of heavy and light particles. 
This is further corroborated in Section~\ref{subsec:lagrange} where we use a Lagrangian refinement method \citep{Hahn_2015} to show that the  error is significantly reduced when the particles are interpolated below the force resolution scale.

\begin{figure*}
    \centering
    \includegraphics[width=\linewidth]{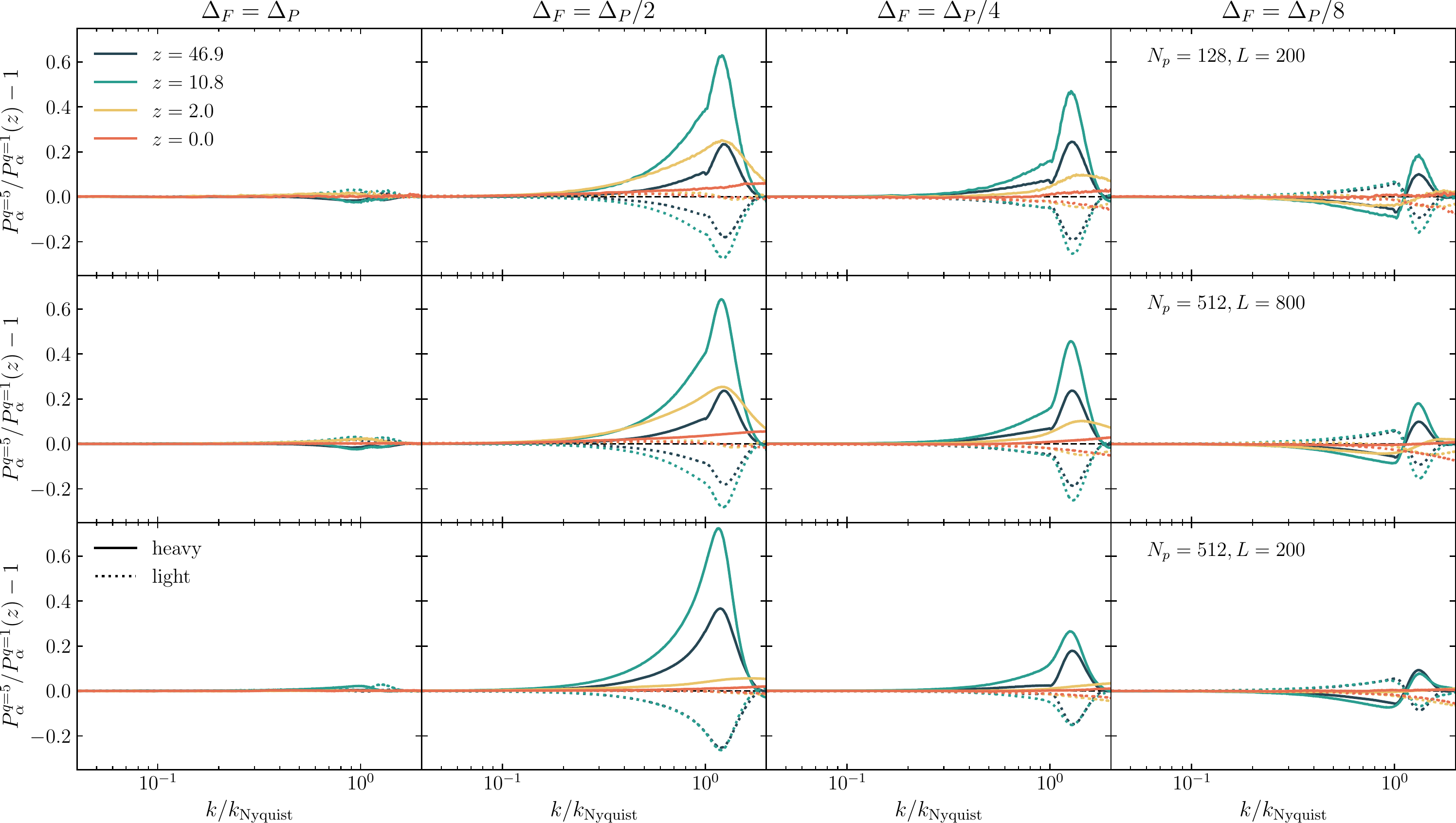}
    \caption{Power spectra of the heavy (solid lines) and light (dotted lines) particle species from an unequal-mass simulation relative to their corresponding power spectra from an equal-mass reference run. Each row corresponds to a fixed simulation box size and particle count, as indicated by the label in the rightmost column. Each column corresponds to a fixed force resolution, as indicated by the label at the top of each column. Colors indicate the redshift of measurement and the x-axis is normalized to the Nyquist frequency, $\knyq = \pi/\DeltaP$, of each simulation.}
    \label{fig:equal_transfer_function_all}
\end{figure*}

\subsection{Asymmetric Sampling}
\label{subsec:sferror} 
We present here results from a suite of 24 simulations that were initialized with two species on offset grids with displacements and velocities set by the total matter transfer function. We consider three simulation box sizes and resolutions: 1) $2\times128^3$ particles in a $\unit{200}{\Mpch}$ box; 2) $2\times512^3$ particles in an $\unit{800}{\Mpch}$ box; 3) $2\times512^3$ particles in a $\unit{200}{\Mpch}$ box. The first two configurations have the same mass resolution which is 64 times coarser than that of the third configuration. For each configuration, we perform three \HACCPM runs that differ in terms of the size of the PM grid with $\NG/\NP = 1$, 2, and 4. In addition, we perform one \HACCTPM run with $\NG = \NP$ and a softening length $r_{\rm soft} = \DeltaP/8$. In other words, for each configuration, we run four simulations where the force resolution varies between $\DeltaF/\DeltaP =$ 1, 1/2, 1/4, and 1/8. Furthermore, for each force resolution, we perform one unequal-mass simulation with mass ratio $q = 5$, as well as a reference simulation with uniform particle masses ($q = 1$). In the latter case, we still use the same cosmological parameters (based on our concordance model) as the $q= 5$ simulation, but simply set all particles to have the same mass. In what follows, we refer to the two particle species in the $q = 5$ runs as ``heavy'' and ``light'' to emphasize that there is no physical distinction between the two groups given that they are initialized using the same transfer function and only gravitational forces are applied.

The results from these simulations are summarized in Fig.~\ref{fig:equal_transfer_function_all}. The first three columns show the \HACCPM runs in ascending order of force resolution while the last column shows the \HACCTPM run. The rows denote the different simulation configurations, as indicated by the label in the final column. 
In each case, we show the power spectra of the heavy and light species from the $q=5$ run relative to the power spectra of the corresponding particle group from the equal-mass reference run. 
By symmetry, the power spectra of the two species in the equal-mass run are identical which means that comparing the $q = 5$ and $q = 1$ runs allows for a direct measure of how the evolution of each species changes when only the masses are varied, with all other parameters held fixed. Note that the absolute power spectra of each species depends on the force resolution of the given simulation, but our focus here is on examining only the relative change that occurs when the particle mass symmetry is broken in the $q = 5$ runs.

When the force resolution is of the same order as the mean interparticle separation (i.e., $\Delta_F = \Delta_P$, as in the first column of Fig.~\ref{fig:equal_transfer_function_all}), the unequal-mass power spectra remain in good agreement with the equal-mass results on all scales and redshifts. However, when the force resolution is made finer than the mean interparticle separation ($\Delta_F < \Delta_P$), the power spectra of the heavy and light species in the $q = 5$ run systematically depart from their corresponding behavior in the $q = 1$ run. In particular, in the two \HACCPM runs in the second and third columns, we see that the heavy (light) species exhibits a systematic enhancement (suppression) of power on small scales around the particle Nyquist frequency. This bias originates at very early times in the simulation, as demonstrated at $z = 47$. The fact that this biased growth is clearly present when the particles are still rather separated from each other suggests that this effect is fundamentally different from particle scattering events such as mass segregation. The bias increases in amplitude as the simulation progresses, peaking around $z = 10$ before subsequently falling in amplitude and being almost entirely erased by $z = 0$. 

Evidently, the qualitative trends are somewhat different in the \HACCTPM run where we instead see an initial suppression (enhancement) of the heavy (light) species for $k \lesssim \knyq$ which then reverses sign at $k \gtrsim \knyq$. The fact that differences arise between the two simulation strategies is perhaps not surprising given that their algorithms differ in their treatment of forces around the particle Nyquist frequency. Nevertheless, the main conclusion remains that the two species exhibit systematically different behavior that operates in opposite directions relative to the equal-mass reference run.

Comparing the first two rows of Fig.~\ref{fig:equal_transfer_function_all} allows us to check the dependence on simulation volume at fixed mass resolution. We find that the results are essentially identical in each column of the first two rows, indicating that the error depends only on the mass resolution of the simulation and is invariant to finite-volume effects. Conversely, comparing the top and bottom rows shows the dependence on mass resolution at fixed volume. At fixed force resolution, the main difference is that the discreteness error is almost entirely erased by $z = 2$ in the bottom row compared to the top row which exhibits biased growth down to $z = 0$. We attribute this to the finer mass and (absolute) force resolution in the bottom row that allows the simulation to resolve smaller scales and capture non-linear growth at earlier times. Non-linear growth aggressively mixes the particle species and seems to wash out the discreteness error that had accumulated earlier in the simulation. A similar argument can be applied to the trend seen when comparing the second and third columns in each row. The increased force resolution in the third column enhances particle mixing which reduces the overall amplitude of the discreteness error at any given redshift.

We can attempt to explain the origin of this discreteness error by considering the nature of the total matter density field seen by the force solver. In the unequal-mass simulations, we are effectively using the masses as a weighting function to up-sample (down-sample) the total matter field at the location of the heavy (light) species. As a result, the total matter field is not smooth below the particle scale and instead oscillates between peaks and troughs in a coherent fashion that follows from the regularity of the offset grids used in the IC. This coherence in the sampling strategy feeds back as a coherent pattern in the gravitational potential at the locations of the heavy and light particles. The result is an asymmetry in the gravitational force calculation that leads to biased growth operating in opposite directions for the two species. If the force resolution is made coarser than the particle scale, then the coherent peaks and troughs average out in the density field, precluding any biased growth. Alternatively, using pre-IC configurations that more adequately mix the two species (as explored in Section~\ref{sec:convergence}), one can diminish the discreteness error. 

\begin{figure}
    \centering
    \includegraphics[width=\columnwidth]{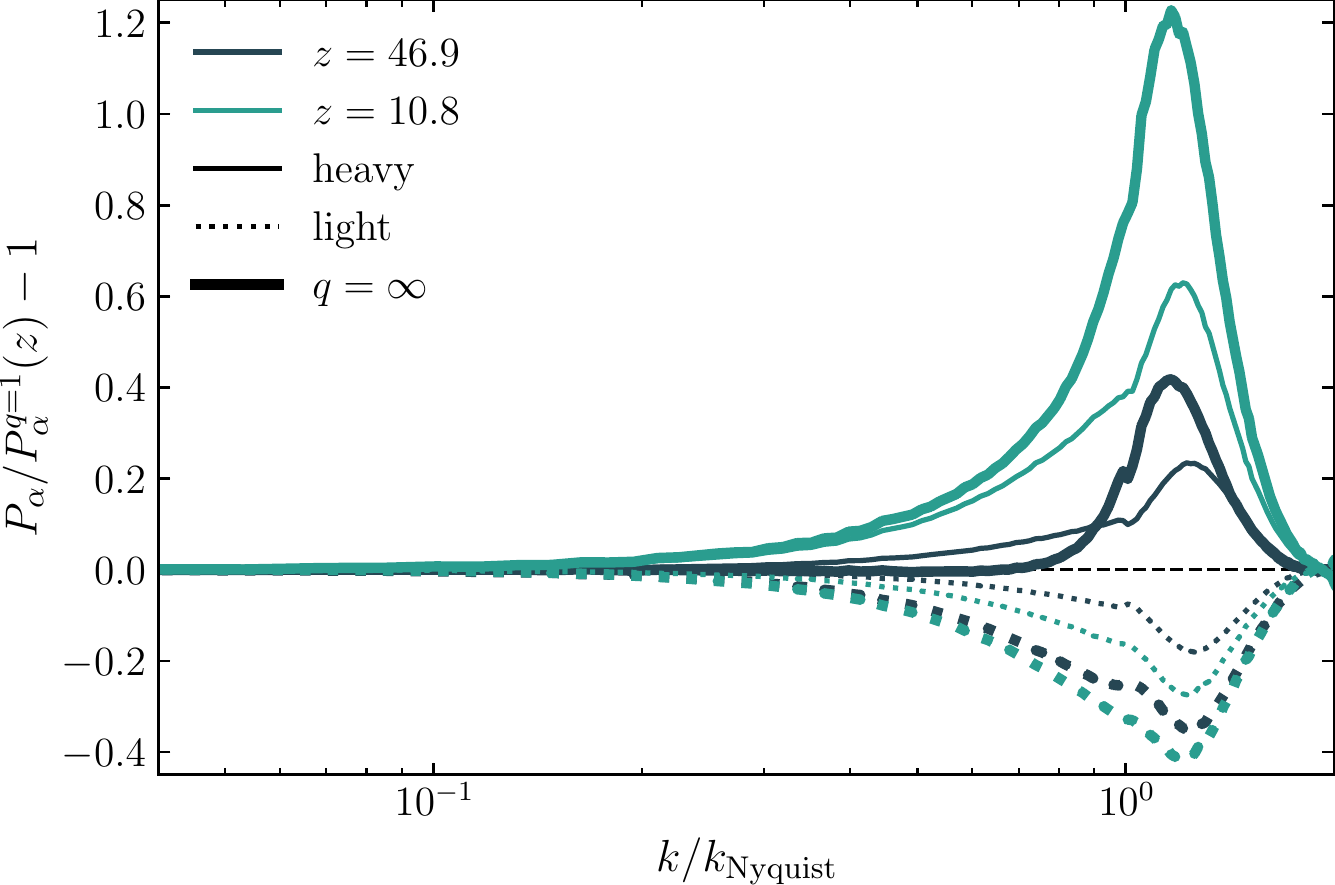}
    \caption{Power spectra of massive (thick solid lines) and passive (thick dotted lines) particles relative to their corresponding power spectra in an equal-mass reference run. For comparison, the thin lines display the corresponding $q = 5$ results shown previously in Fig.~\ref{fig:equal_transfer_function_all}. Each run evolves $2\times128^3$ particles in a $\unit{200}{\Mpch}$ box using \HACCPM with $\DeltaF = \DeltaP/2$.}
    \label{fig:equal_transfer_function_passive}
\end{figure}

We can explore this idea further by considering the asymptotic limit of the tests above where the mass ratio is set to infinity ($q=\infty$). In this idealized scenario, the particle mass of one species is set to zero meaning that this species acts as a passive tracer of the gravitational potential sourced by the massive species. The result is shown in Fig.~\ref{fig:equal_transfer_function_passive} where we use the first simulation configuration described above ($2\times128^3$ particles in a $\unit{200}{\Mpch}$ box) performed using \HACCPM with a PM mesh so that $\DeltaF = \DeltaP/2$. As before, we normalize the results to the equal-mass reference simulation evolved under identical settings. For comparison, we also display the $q = 5$ run shown previously in the second column of the top row in Fig.~\ref{fig:equal_transfer_function_all}. 

We find qualitatively similar results to the $q = 5$ simulations. Namely, we see that the passive species displays heavily suppressed growth on small scales relative to the massive species. We can attribute this to the systematic offset of $\DeltaP/2$ between the locations at which the massive species sources the density field and the locations at which the passive species samples the gradient of the gravitational potential. By symmetry, we would instead  find identical growth between the two species if the passive particles were initialized on top of the massive particles. Comparing to the $q = 5$ case shows how this scenario changes when we transfer some of the mass from the massive to the passive species. The result is that the light particles still exhibit suppressed growth relative to the heavy species, but with a reduced amplitude compared to the $q = \infty$ case. Finally, symmetry dictates that the scenario changes in the $q = 1$ case with both species growing in an identical manner.

To summarize, these tests demonstrate that using unequal particle masses to coherently up- and down-sample the total matter field leads to a corresponding growth bias operating in opposing directions for the two species. Relative to the symmetric equal-mass case, we find that the heavy (light) particles acquire enhanced (suppressed) growth on scales around the Nyquist frequency when evolved with a PM solver. The situation changes slightly when pairwise forces are used on small scales, but the main conclusion of asymmetric biased growth remains. This error stems from the fact that the force solver sees a total matter density field that is coherently discretized into unmixed components. Eventually, the error stops growing when the particle species become sufficiently mixed and the coherence in the density field is smoothed out. Subsequent non-linear growth tends to overtake the initial accumulation of this error. 

\subsection{Lagrangian Refinement}
\label{subsec:lagrange}

To further illustrate that the biased growth arises from an unmixed sampling of the total matter density field, we utilize the Lagrangian refinement method of \citet{Hahn_2015} to smooth the particle discretization for a fixed force resolution. Each pre-IC grid cell is effectively replaced by $4^3=64$ refined particles uniformly distributed within the Lagrangian cube. We perform this upsampling on-the-fly during each force calculation step, without having to run the entire simulation at a higher particle resolution. This is achieved through a continuous mapping between Lagrangian space and the phase-space distribution, $\textbf{q} \mapsto (\textbf{x}_q(t), \textbf{v}_q(t)$), where we utilize a tri-linear interpolation map during the CIC deposition step in our integrator. While we do not expect the refining procedure to resolve the numerical solution at higher fidelity, we do expect it to mitigate discretization errors as the density sampling is now being smoothed.

\begin{figure}
    \centering
    \includegraphics[width=\columnwidth]{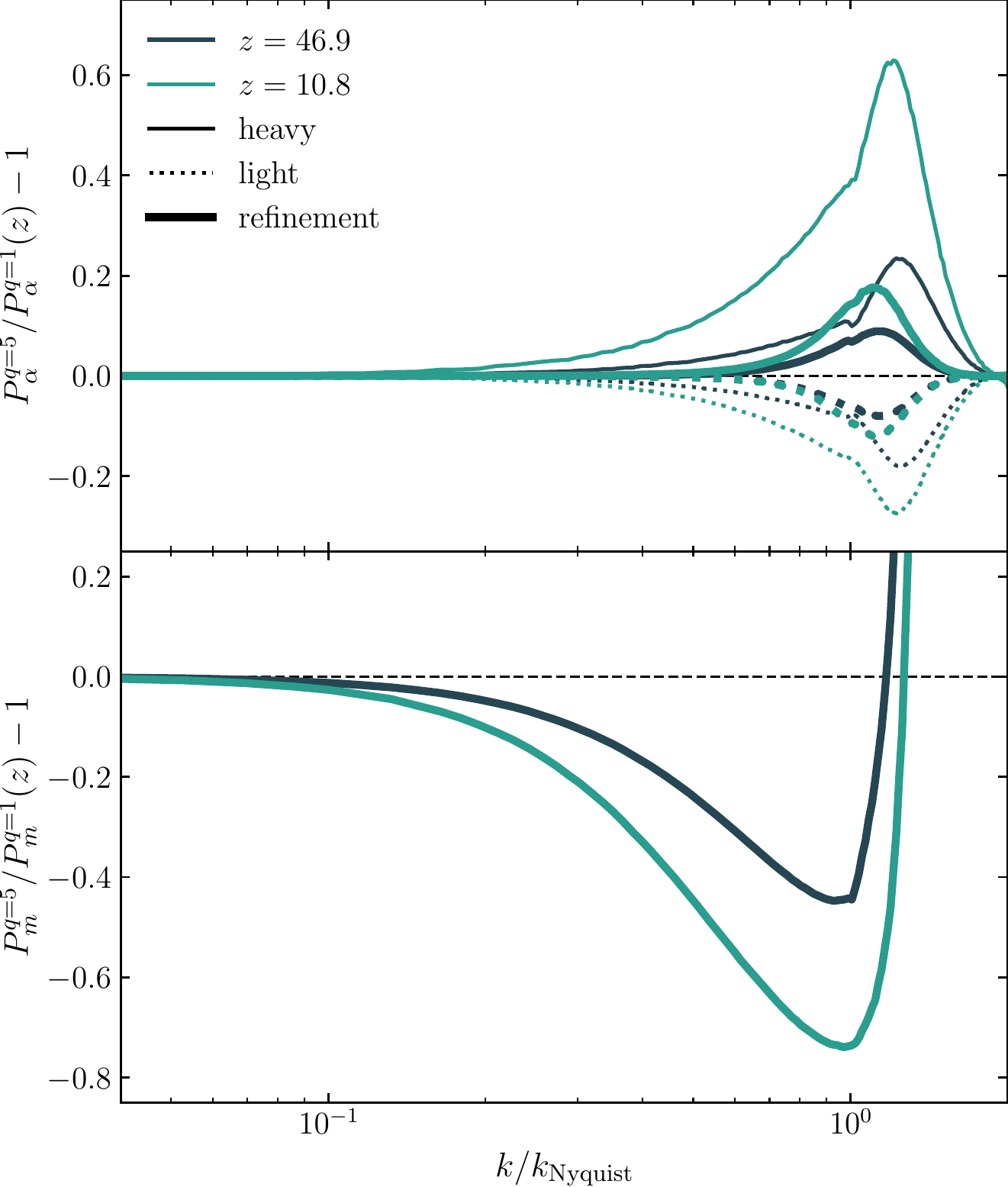}
    \caption{Power spectra of the heavy (solid lines) and light (dotted lines) particles relative to their corresponding power spectra in an equal-mass reference run. Thick and thin lines correspond to runs with and without Lagrangian refinement, respectively. Each run evolves $2\times128^3$ particles in a $\unit{200}{\Mpch}$ box using \HACCPM with $\DeltaF = \DeltaP/2$. The bottom panel compares the total matter power spectrum from the refined run relative to the $q = 1$ result.}
    \label{fig:equal_transfer_function_refined}
\end{figure}

 In Fig.~\ref{fig:equal_transfer_function_refined} we compare power spectra from one of the $q = 5$ simulations performed both with and without the Lagrangian refinement method. Each simulation evolved $2\times128^3$ particles in a $\unit{200}{\Mpch}$ box using \HACCPM with $\DeltaF = \DeltaP/2$. In the top panel, we present the power spectrum of the heavy and light species from the two $q = 5$ runs relative to the $q = 1$ reference simulation; the bottom panel compares the total matter power spectra. We see that the Lagrangian refinement significantly reduces the level of biased growth in both the heavy and light species. In contrast, the total matter power spectrum is heavily suppressed relative to the reference run.

The Lagrangian refinement run differs in only the fact that the particle density field is interpolated to a scale 4 times finer than the mean interparticle separation. The result is a significant reduction in the biased growth on small scales that is achieved at fixed force resolution without increasing the actual number of simulation particles. This result reinforces the notion that the biased growth is attributed to the force solver resolving an unmixed particle distribution. We note that the Lagrangian refinement method is accurate only as long as the phase-space submanifold is well-sampled by the $N$-body particles \citep[e.g., at early times in WDM simulations, see][]{Hahn_2015}. Once the particle sampling is no longer sufficient to trace the folding of the dark matter sheet in phase-space, the now unreliable mapping from Lagrangian to Eulerian space leads to a suppression of growth near the Nyquist frequency, as shown in the bottom panel of Fig.~\ref{fig:equal_transfer_function_refined}. Given this drawback, we do not recommended using Lagrangian refinement to address discreteness errors, but rather present it here as an illustrative example of how the biased growth is fundamentally sourced from the discrete nature of the density field in multi-species simulations. 


\section{Two-fluid Simulations}
\label{sec:unequaltransfer}
We now investigate the consequences of initializing simulations with individual density and velocity transfer functions for CDM and baryon particles. This is the more physically consistent approach for cosmological simulations despite being rarely used in practice. In order to separately distinguish numerical artifacts seeded by particle mass discrepancies (as investigated in Section~\ref{sec:unequalmass}), we begin by studying uniform mass ($q=1$) simulations in Section~\ref{subsec:uniformcosmo}. In this case, we find that both species exhibit biased growth on large scales reinforcing the notion that the error is sourced from the unmixed discretization of the total matter field into its CDM and baryon components as opposed to arising purely from having unequal particle masses. We then further study unequal-mass ($q=5$) simulations in Section~\ref{subsec:concosmo}, which are found to behave in a similar fashion as the equal-mass case, but with the discreteness error shifted mostly into the lighter baryon component. 

\begin{figure*}
    \centering
    \includegraphics[width=\linewidth]{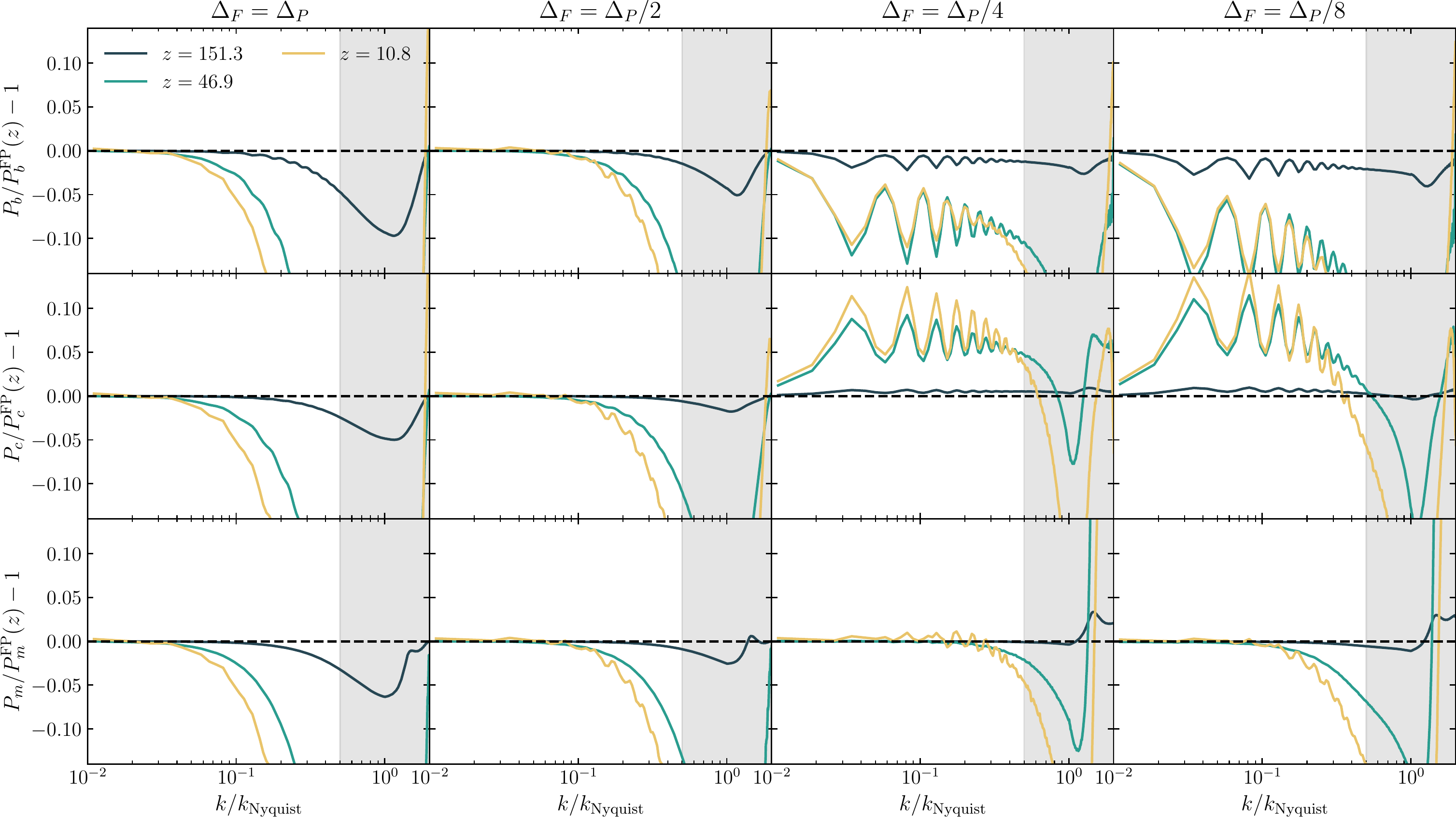}
    \caption{Power spectra of baryons (top), CDM (middle), and total matter (bottom) from two-fluid simulations performed using an equal density cosmology ($\Omega_c = \Omega_b$) so that $q = 1$. The results are shown in comparison to a reference \HACCPM run evolved under the FP approximation with $\DeltaF = \DeltaP$. Each simulation evolved $2\times256^3$ particles in a $\unit{512}{\Mpch}$ box. Each column corresponds to a different run, with the first three being \HACCPM runs with $\DeltaF/\DeltaP = 1$, $1/2$, and $1/4$, respectively, while the last column corresponds to a \HACCTPM run with $\DeltaF/\DeltaP = 1/8$. Attention should be restricted to the non-shaded regions in each panel since smaller scales are heavily impacted by the CIC interpolation function used in the simulations but absent in the FP model.}
    \label{fig:unequal_transfer_function_q1}
\end{figure*}

\begin{figure}
    \centering
    \includegraphics[width=\columnwidth]{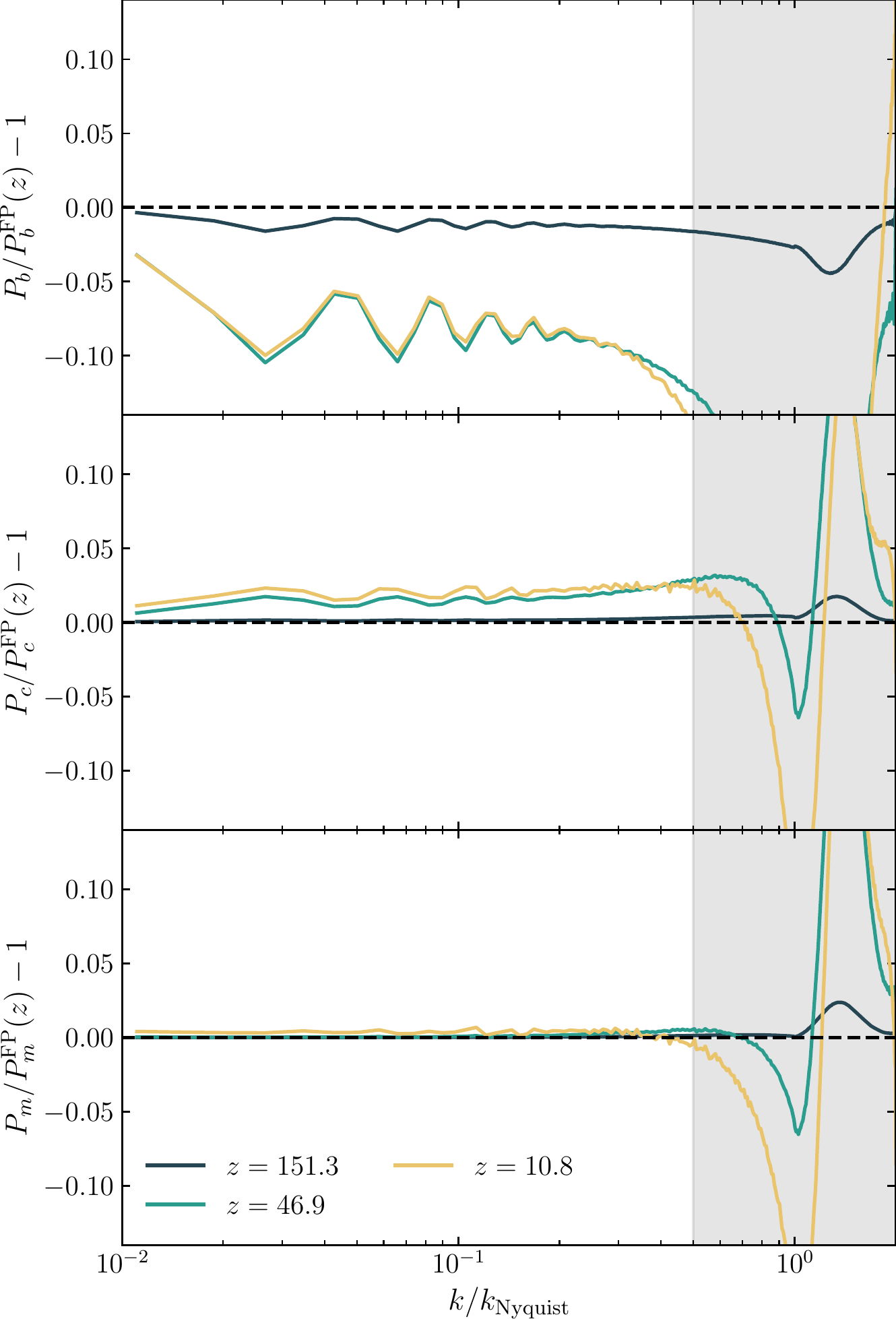}
    \caption{Power spectra of baryons (top), CDM (middle), and total matter (bottom) from a two-fluid simulation performed using the concordance cosmology ($\Omega_c = 5\Omega_b$) so that $q = 5$. The simulation evolved $2\times256^3$ particles in a $\unit{512}{\Mpch}$ box using \HACCPM with $\DeltaF/\DeltaP = 1/4$ and is compared to a baseline FP run. Attention should be restricted to the non-shaded regions in each panel since smaller scales are heavily impacted by the CIC interpolation function used in the simulations but absent in the FP model.}
    \label{fig:unequal_transfer_function_q5}
\end{figure}

\subsection{Balanced Cosmology}
\label{subsec:uniformcosmo}

We begin with an analysis of a set of simulations that employ an equal density cosmology with $\Omega_c = \Omega_b$ so that the resulting CDM and baryon particle masses are equal. A total of four simulations were run, with each evolving $2\times256^3$ particles in a $\unit{512}{\Mpch}$ box. As before, we performed three \HACCPM runs that vary the PM mesh size so that $\DeltaF/\DeltaP = 1$, 1/2, and 1/4, as well as one \HACCTPM run where the PM mesh is fixed with $\NG = \NP$ and the softening length is set so that $\DeltaF/\DeltaP = 1/8$.  The results are shown in Fig.~\ref{fig:unequal_transfer_function_q1} with each column denoting an individual run while the rows separately show the baryon, CDM, and total matter power spectra, respectively. 

The power spectra of each simulation is shown relative to a reference run evolved under the FP approximation. The reference run was performed using \HACCPM with $\DeltaF = \DeltaP$ and, as described in Section~\ref{sec:method}, evolved particles under the LT total matter field at each time step. We used the FP run as a baseline as opposed to simply comparing against LT since we found that the white noise profiles of the individual CDM and baryon power spectra become mixed when evolved together, obfuscating the signatures that we are looking for on large scales. While the FP method achieves the correct linear growth of each species on large scales (see Fig.~\ref{fig:frozen_potential}), it does omit non-linear terms meaning that it is a reliable baseline only at high redshift. In addition, the FP method differs from the actual simulations in that it does not use any interpolation function to compute the density field. As such, we cannot make an apples-to-apples comparison between the simulations and the FP baseline on small scales. Neither of these issues are important for our application, however, since we are mainly focused on large scales and early times.

The results in Fig.~\ref{fig:unequal_transfer_function_q1} are analogous to those found earlier for the single-fluid simulations in Fig.~\ref{fig:equal_transfer_function_all}. In particular, we find that the CDM, baryon, and total matter power spectra match the growth of the FP method on large scales for the runs in the leftmost two columns where $\DeltaF/\DeltaP \geq 1/2$. Conversely, we observe a striking contrast in the growth of the CDM and baryon components in the rightmost two columns for which $\DeltaF/\DeltaP \leq 1/4$. Here we see that the CDM (baryons) grow systematically fast (slow) relative to the baseline run. Unlike the single-fluid simulations, the biased growth in the two species is clearly visible out to the largest scales resolved in the simulation. Moreover, we see prominent wiggles in both the CDM and baryon curves on large scales. This reflects a modulation in the BAO signal that results from a delayed transfer of BAO power from baryons to CDM due to the biased growth of each species. Given that the errors extend to large scales, we do not observe any significant differences between the \HACCPM and \HACCTPM runs which differ only in their treatment of small-scale forces. Interestingly, we find that the biases of the individual components in each simulation cancel out in the total matter field which is seen to match the large-scale growth of the reference run.

We argue that the errors seen in the single- and two-fluid simulations are manifestations of the same problem; namely, the unmixed discretization of the total matter field. In the single-fluid case, the confinement of the error around the Nyquist mode reflects the fact that the discretization occurs only at the particle scale. In the two-fluid case, the error extends to the much larger scales where the individual transfer functions differ from that of the total matter. Indeed, in simulations with much larger boxes, we have found that the bias does not materialize on scales where the CDM and baryon transfer functions converge ($k \lesssim \unit{10^{-2}}{\invMpch}$). Importantly, the bias occurs even in simulations with uniform particle masses. This confirms that the large-scale error seen in two-fluid simulations is fundamentally sourced by the discretization of the total matter field and not from particle masses per se. We reinforce this interpretation using a plane wave case-study in Appendix~\ref{app:planewave}, exploring discreteness effects in an idealized setting. 

\subsection{Concordance Cosmology}
\label{subsec:concosmo}

The previous subsection served as an illustrative example of how large-scale discreteness errors still occur in simulations with uniform particle mass. We switch attention here to the more realistic concordance cosmology which leads to unequal-mass ($q=5$) CDM and baryon particles. This is similar to the mass ratio used in previous works (\citealt{angulo2013}; \citetalias{Bird_2020}; \citetalias{Hahn_2020}) that have also studied discreteness errors in two-fluid simulations. 

For this test, we ran a single \HACCPM simulation that evolved $2\times256^3$ particles in a $\unit{512}{\Mpch}$ box with $\DeltaF = \DeltaP/4$. We also performed a baseline run where the particles are evolved under the FP approximation with $\DeltaF = \DeltaP$. We compare the power spectra from these two runs in Fig.~\ref{fig:unequal_transfer_function_q5}. This plot can also be compared to the third column of Fig.~\ref{fig:unequal_transfer_function_q1} where the same set of parameters (apart from the cosmology) are used. We observe here the same qualitative features as seen previously for the equal density cosmology. More specifically, the large-scale offsets and BAO modulation are present in both the baryon and CDM power spectra. The main difference, however, is that the bias in the CDM power spectra is much more subdued in the concordance cosmology. We attribute this to the fact that the CDM now dominates the total matter field meaning that the error is preferentially shifted into the subdominant baryon component. We again find that the large-scale growth of the total matter field aligns well with the reference run.

These results are consistent with the notion that errors arise from the unmixed discretization of the total matter field. In this case, the total matter field is split into particles with asymmetries in both mass and power spectra. The results of Section~\ref{sec:unequalmass} suggest that the particle masses impart errors on small scales while the findings in Section~\ref{subsec:uniformcosmo} confirm that the disparate density fields are responsible for driving errors to large scales. In the next section, we examine various methodologies that can be used to address these issues.


\section{Evaluation of Mitigation Strategies}
\label{sec:convergence}

We conclude our analysis with an evaluation of several approaches to address the large-scale discreteness errors seen in two-fluid simulations. For the remainder of this section, we utilize target simulations evolving $2\times256^3$ particles in $\unit{64}{\Mpch}$ boxes. This setup was chosen so that the mitigation strategies are evaluated using a mass resolution that is more representative of production-scale simulations used to generate synthetic skies. With our concordance cosmology, the CDM and baryon particle masses are $m_c = \unit{1.1\times10^9}{\massh}$ and $m_b = \unit{2.2\times10^8}{\massh}$, respectively. The simulations are performed using either \HACCPM with an $\NG = 1024$ PM mesh ($\DeltaF = \DeltaP/4$) or \HACCTPM with $r_{\rm soft} = \DeltaP/20$. We have also performed a reference \HACCPM run that was chosen to have the same force resolution as the \HACCPM target runs but containing $2\times1024^3$ particles ($\DeltaF = \DeltaP$) so that large-scale discreteness errors are absent. The reference run was initialized using the same white noise field as the target runs. 

In Section~\ref{subsec:preICs}, we investigate two initial condition approaches recently proposed by \citetalias{Bird_2020} and \citetalias{Hahn_2020}, and evaluate how well they reduce the offset power bias. We further analyze simulations evolved at lower starting redshifts in Section~\ref{subsec:zl}, where the sampling distribution is more mixed. Lastly, in Sections~\ref{subsec:mitfp} and \ref{subsec:sfs} we discuss new methods that can mitigate the discreteness errors while using standard grid ICs.  

\subsection{Initial Condition Configurations}
\label{subsec:preICs}

In this section, we investigate the effects on the power spectrum when we start simulations with different ICs. Three separate initialization layouts at $z_i=200$ are considered here as follows: offset grid, mixed grid plus glass \citepalias{Bird_2020}, and mass perturbed \citepalias{Hahn_2020} ICs.
Briefly, the configuration proposed by \citetalias{Bird_2020} is designed to efficiently mix the two species by initializing the CDM particles on a grid while the baryons use a glass distribution \citep{glass_note}. The method described in \citetalias{Hahn_2020} utilizes offset grids by displacing particles with the total matter transfer function, and therefore, avoids issues with unmixed discretization of separate fluids (with the additional benefit of optionally initializing to higher order LPT).
The individual species transfer functions are separately encoded using mass and velocity perturbations. 
We use the publicly available MP-GenIC \citep{yu_feng_2018_1451799} code to generate the \citetalias{Bird_2020} ICs and \MUSIC\footnote{Available from \url{https://bitbucket.org/ohahn/monofonic}} for the \citetalias{Hahn_2020} runs.

\begin{figure}
    \centering
    \includegraphics[width=\columnwidth]{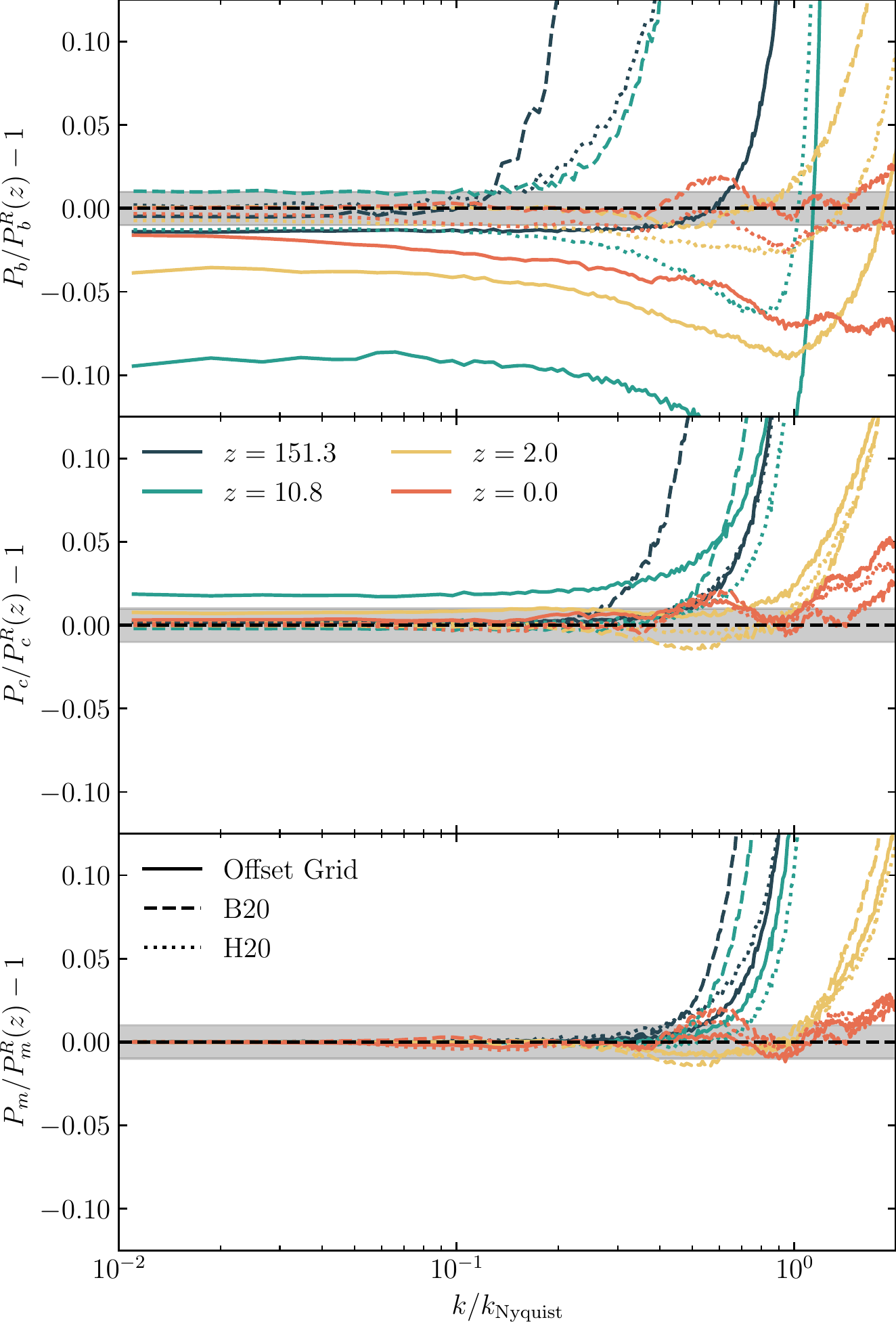}
    \caption{Power spectra comparison of \HACCPM simulations starting with three different ICs (labelled with different line-styles) and the reference simulation. The simulations are performed with $2\times256^3$ particles in a $64 \ h^{-1}$Mpc box. The panels from top to bottom present the power spectrum ratios for baryons, CDM, and total matter at a fixed PM force resolution of $\DeltaF = 1/4 \DeltaP$. The horizontal shaded gray band in each panel corresponds to deviations of $\pm1\%$.} 
    \label{fig:convergence_3ics}
\end{figure}

The results of three \HACCPM simulations initialized with the methods described above are presented in Fig.~\ref{fig:convergence_3ics}. From top to bottom, the panels show the power spectrum ratios between the simulations and the reference run for baryons, CDM, and the total matter. For the offset grid ICs, we find the previously discussed deviations (see Section~\ref{sec:unequaltransfer}) in the baryon and CDM power spectrum at all $k$, peaking at $z\sim10$  with approximately $10\%$ suppression for baryons and $2\%$ enhancement for CDM. At later times, the large-scale offsets decrease as non-linear growth starts to govern the evolution at high $k$. Note that the discreteness errors only affect the growth of the individual species; the total matter power agrees with the reference simulation at low $k$.

In comparison, the grid$+$glass runs show significantly improved power spectra results, where the large-scale differences are within $2\%$ for baryons and less than $1\%$ for CDM at all redshifts.
This reflects the pseudo-random nature of the glass which prevents coherence in the discretization of the total matter field into its two components. The drawback of the glass, however, is the introduction of noise on small scales, manifesting as an excess of power at medium and high $k$ in the individual and total matter distributions. We note that \citetalias{Bird_2020} has also shown that undersampling the baryonic fluid to maintain equal mass between the species results in similar improvements to a mixed grid$+$glass IC. We have confirmed this to be true using an offset grid simulation containing $256^3$ CDM and $150^3$ baryon particles. In accordance with our previous results, this improvement likely arises from symmetry breaking due to the irregular grid separation between species and not from having the same particle masses. Given that equal particle sampling of both fluids is the most common method of discretization, we focused on that configuration throughout this study. 

Turning to the results of the mass-perturbed simulations, we find that the power spectra displays
roughly $3\%$ suppression for baryons at $z=151.3$ and less than $1\%$ enhancement for CDM at all four redshifts. The baryon suppression quickly drops to the sub-percent level for $z \leq 10$. The success of the mass-perturbed method can be attributed to the fact that the individual transfer functions are absorbed into the masses as opposed to the initial displacements. In this way, the force solver sees a particle distribution that is consistent with the total matter field on large scales, but with small mass perturbations at the particle scale. Over time, the separate initial velocity fields will lead the particles to ``dephase'' meaning that the original issue causing the large scale offsets will resurface (i.e., see the small bias in baryon power at $z = 10.8$ that is later reduced by non-linear growth). For this reason, the mass-perturbed method is more aptly suited for lower redshift initializations, as we show in the next section. 

\subsection{Low Redshift Initialization}
\label{subsec:zl}

A direct approach to address errors incurred by evolving unmixed distributions is achieved by initializing the simulation at lower redshifts, perturbing the particles further from their initial grid positions.
A major concern about late simulation starts is the requirement of higher order LPT to accurately capture the curved trajectories of the initial displacement. 
In the single-fluid case, studies have illustrated at least second order perturbation theory (2LPT) is desired for initializing accurate N-body simulations at $z_i=50$ \citep[e.g.][]{Schneider_2016,Michaux_2020}. However, there is not yet a consistent method to compute higher order LPT for two-fluid simulations. The procedure outlined in \citetalias{Hahn_2020} does utilize higher order LPT for the combined baryon and CDM distribution, albeit using an initial relative velocity field, $v_{\rm bc}$, calculated only to first order. We will use both mass-perturbed (initialized with first and second order LPT) and standard grid configurations to study low redshift initial conditions below. 

\begin{figure}
    \centering
    \includegraphics[width=\columnwidth]{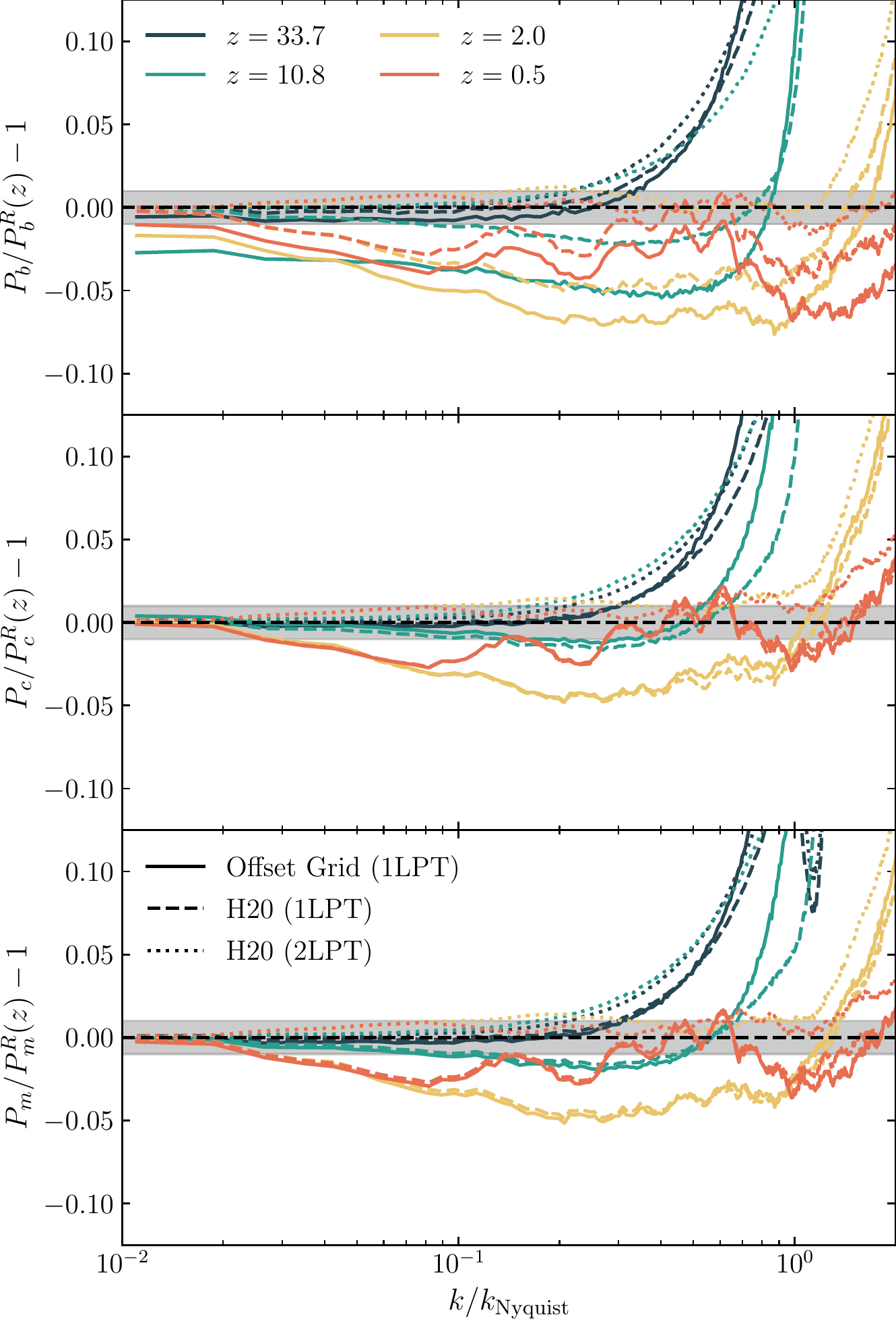}
    \caption{Power spectra of \HACCPM simulations initialized at $z_i=46.9$ with the offset grid ICs (solid lines), and mass-perturbed \citetalias{Hahn_2020} ICs with first order (dashed lines) and second order (dotted lines) LPT, compared to the reference simulation. The horizontal shaded gray band in each panel corresponds to deviations of $\pm1\%$.}
    \label{fig:convergence_lowz}
\end{figure}

In Fig.~\ref{fig:convergence_lowz}, we show the power spectrum results from \HACCPM simulations initialized with three different methods at $z_i=46.9$: a standard grid setup with first order LPT and mass-perturbed runs with both first and second order LPT.
Starting with the standard grid runs, we notice that although the large-scale offsets in the baryon and CDM power spectrum are still present, they are significantly reduced compared to the run initialized at $z_i=200$ (Fig.~\ref{fig:convergence_3ics}). More specifically, the power suppression for baryons on large scales lowers from approximately 10\% to 3\% at $z=10.8$. 
Next, the simulation initialized using the mass-perturbed method of \citetalias{Hahn_2020} with 1LPT further improves the large-scale behavior of the individual species power, matching the high resolution reference simulation at the lowest $k$ modes. While the individual power spectra for the 1LPT standard grid and mass-perturbed runs differ on large scales, their total matter power remain in broadly good agreement on the full set of scales and redshifts. On mid to small scales, however, there is a clear lack of total matter power compared to the reference simulation. This discrepancy is significantly improved using the mass-perturbed 2LPT initial condition which displays sub-percent agreement with the reference run at $z = 0.5$.
Hence, the mass-perturbed 2LPT run not only removes the large-scale offsets, but also more closely matches the reference simulation at intermediate scales. 

Thus, we have confirmed that lower redshift starts can reduce the biased growth in power on large scales. Moreover, the mass perturbation approach further addresses the necessity for higher order LPT initial conditions, in an improved manner over simple grid runs. Of course, if one wishes to study high-redshift structure formation (i.e., first galaxy formation) for which early start simulations are needed, consideration of the necessary particle sampling to overcome these effects is required.  
We exclusively utilize power spectrum diagnostics in this paper but there are in principle other metrics (e.g., bispectrum, halo mass function) that could be impacted by initializing the simulation at lower redshifts. Further studies regarding this topic are currently in progress. 

\subsection{Frozen Potential Approach}
\label{subsec:mitfp}

We present here an alternative method for starting offset grid simulations at low redshift while still maintaining a first-order initialization. The idea is to construct 1LPT initial conditions in the usual manner at $z_i=200$, followed with subsequent evolution to some later handover redshift, $z_h$, under the FP approximation (see Section~\ref{subsec:fp}). The remaining evolution is completed with the full force solver. In this way, the FP method can be thought of as an extension to the initial conditions. Of course, care must be taken on the choice of $z_h$ since a value too high will still suffer from discreteness errors while a value too low will delay non-linear growth. In general, the compromise between these limits will depend on the resolution of the simulation.

\begin{figure}
    \centering
    \includegraphics[width=\columnwidth]{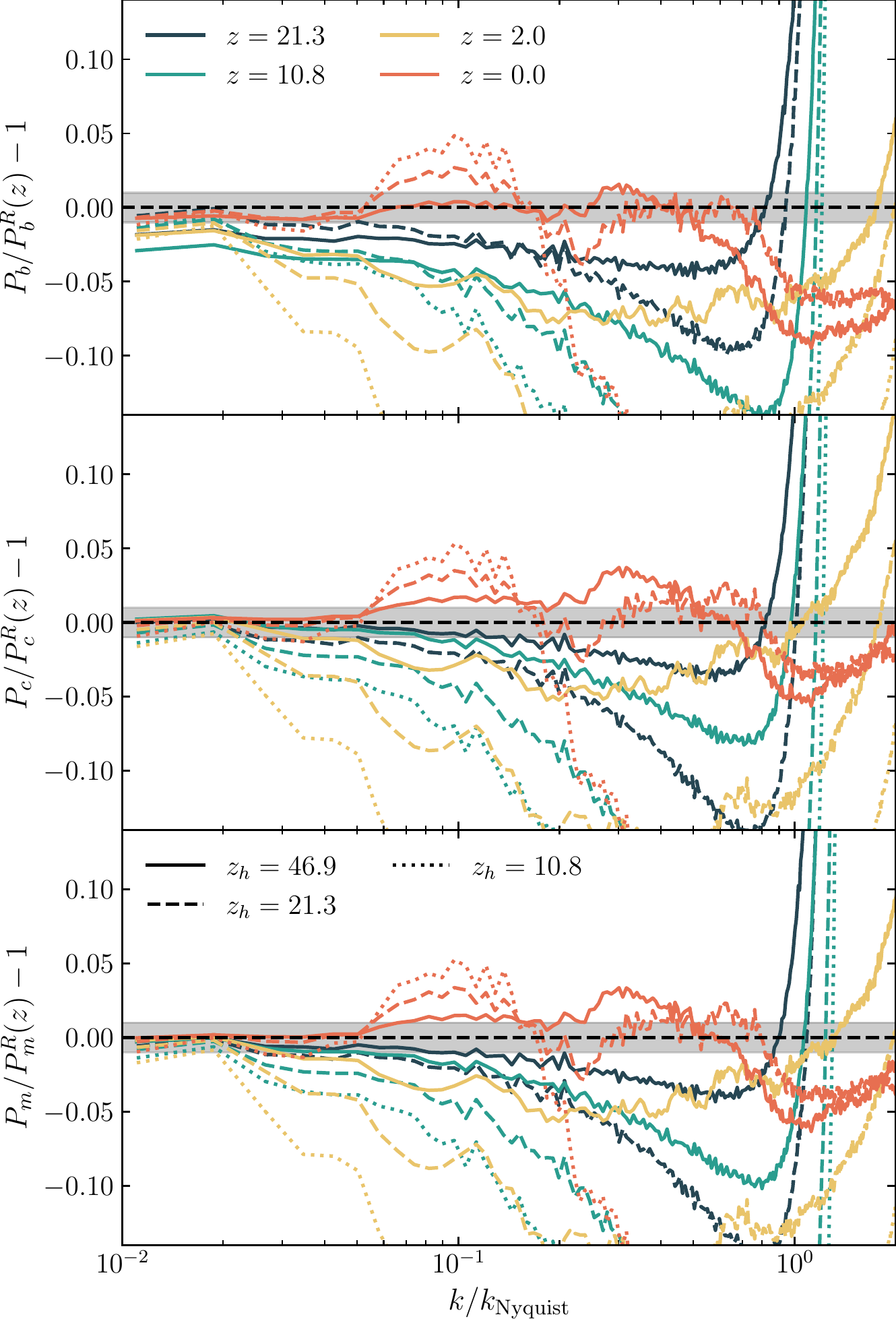}
    \caption{Power spectra of the FP simulations with three different handover redshifts $z_h$ compared to the reference simulation. The FP simulations are evolved under the ``frozen'' potential to the handover redshift $z_h$, after which the regular \HACCPM algorithm with $\DeltaF = 1/4 \DeltaP$ is applied. The horizontal shaded gray band in each panel corresponds to deviations of $\pm1\%$.}
    \label{fig:convergence_frozen}
\end{figure}

To explore this further, we perform three runs with our target simulation where the FP method is used until $z_h=46.9$, $21.3$, and $10.8$. In each case, \HACCPM is used to evolve the particles from $z_h$ to $z = 0$. The results of these tests are organized in Fig.~\ref{fig:convergence_frozen}.
The simulation with the highest handover redshift, $z_h=46.9$, shows approximately $3\%$ large-scale suppression in baryon power spectrum at $z=10.8$ and no significant large-scale offsets in CDM and total matter power spectrum at any redshift. This is a significant improvement compared to the $10\%$ suppression for baryons and $2\%$ enhancement for CDM shown in Fig.~\ref{fig:convergence_3ics} for the default target run with $z_i = 200$. On intermediate scales, however, there is a clear suppression in both baryon and CDM power for $z \geq 2$. This power suppression is replaced at $z = 0$ with random noise at the few-percent level. At this point, the simulation volume is almost entirely non-linear which appears to mostly erase the earlier systematic power suppression.
Lowering the handover redshift to $z_h=21.3$, the large-scale deviation in the baryon power spectrum further reduces to less than $2\%$. Unsurprisingly, the power suppression on intermediate scales has become more pronounced, though this is once again mostly erased by $z = 0$. Finally, the simulation with $z_h=10.8$ shows significant power suppression at all times indicating that such a low handover redshift is not appropriate for capturing non-linear growth.

In summary, we find that using the FP approximation to evolve particles from first-order ICs at high redshift to later times can significantly reduce large-scale biases resulting from insufficient particle mixing. The challenge is that the FP method is limited by both its assumption of linear growth as well as a coarser force resolution due to the omission of short-range forces. In general, a careful analysis of the appropriate choice of $z_h$ would need to be performed for a given mass resolution, in addition to consideration of other metrics outside of the power spectrum. 

\subsection{Softening Interspecies Interactions}
\label{subsec:sfs}

Our final mitigation strategy for reducing the large-scale power bias is connected to the observation that these errors occur when the force resolution is finer than the interparticle separation. In particular, errors are most pronounced in undersampled regions where the unmixed initial grid configuration of each species maximally impacts the evolution. In response, we explore a modification to our short-range gravity solver which adaptively turns off or softens interspecies interactions in low-density regions.

Within \HACC, the short-range solver operates on neighboring chaining mesh (CM) cells of size $l_{\rm CM} = 4\DeltaG$ (see Section~\ref{subsec:sims}). 
Utilizing the same spatial decomposition, we implement a particle number per cell threshold, $N_t$, below which the short-range cross-species interaction forces are turned off or softened for any particle within the cell. These include pairwise force contributions from cell neighbors. 
This modification only affects low density regions within the simulation, with the intention to not suppress non-linear growth in clustered regions. Note that for locations of the simulation at average background density we expect $N_{\rm CM} = 2\times 4^3=128$ particles per cell. We found for our simulation resolution that a value of $N_t=160$, corresponding to an overdensity $\delta_\mathrm{CM} = 0.25$, to be a good compromise between alleviating the discreteness errors while preserving the growth of non-linear structures and we adopt this threshold in the following.

We implement three different variations of the short-range force operator: \SFSSOFTEN, \SFSOFF, and \SFSBOOST. \SFSSOFTEN is motivated by previous mitigation attempts using adaptive gravitational softening lengths \citep[e.g.,][]{angulo2013,Villaescusa_Navarro_2018} and the observation that the discreteness errors disappear when using \HACCPM with $\DeltaF=\DeltaP$ (see Section~\ref{sec:unequaltransfer}). In cells below the particle threshold $N_{\rm CM}<N_t$, we increase the softening length from $r_{\rm soft} = \DeltaP/20$ to $r_{\rm soft} = \DeltaP$ for cross-species interactions within the cell and its neighbors, effectively reverting the gravity solver to \HACCPM with $\DeltaF=\DeltaP$ for those interactions.
\SFSOFF extends that modification by completely turning off cross-species short-range forces, thereby isolating the effects of interspecies interactions on structure formation.
Finally, in an attempt to compensate for the suppression of growth on small scales, we implement \SFSBOOST, which boosts the particle masses by an amount $\Omega_m/\Omega_\alpha$ in same-species interactions when cross-species forces are turned off to match the total matter density. Note that the PM force calculation is unmodified in each of these cases meaning that cross-species interactions are still present on large scales.

\begin{figure}
    \centering
    \includegraphics[width=\columnwidth]{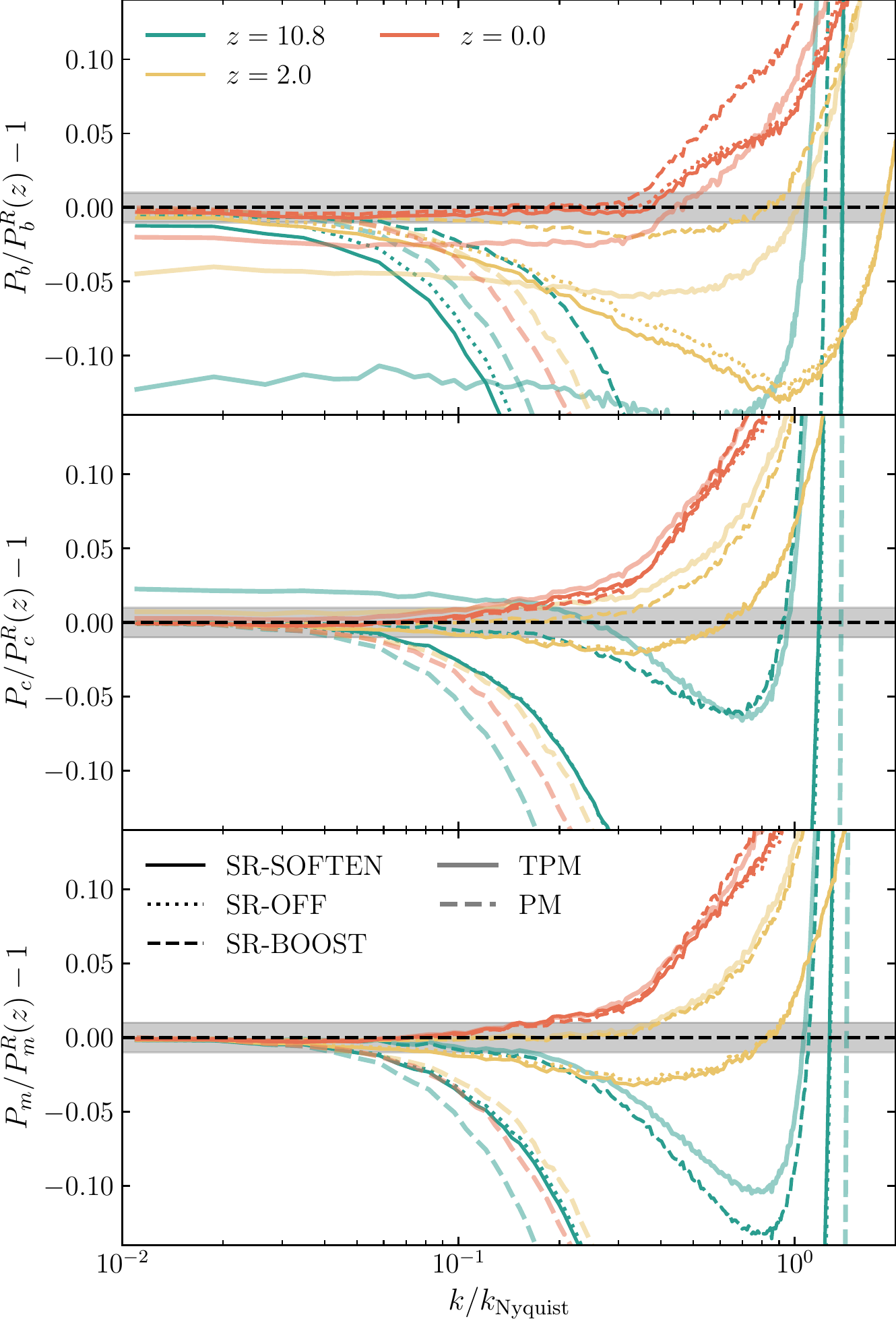}
    \caption{Power spectra of the simulation runs with the \SFSSOFTEN, \SFSOFF, and \SFSBOOST modification compared to the high-resolution reference simulation. We also include the \HACCPM and \HACCTPM simulations for reference. The simulations are run with the same softening length as \HACCTPM, but additionally incorporate a density threshold $N_{\rm CM} = 160$ below which the interspecies interactions are softened with $r_{\rm soft} = \DeltaP$ (\SFSSOFTEN) or turned off (\SFSOFF, \SFSBOOST). In the last implementation, same-species interactions are additionally boosted to compensate for the missing short-range forces. The horizontal shaded gray band in each panel corresponds to deviations of $\pm1\%$.}
    \label{fig:convergence_cms}
\end{figure}

The power spectrum ratios between the simulations with modified interspecies forces and the high-resolution \HACCPM reference simulation are shown in Fig.~\ref{fig:convergence_cms}. We also include the \HACCTPM simulation  with constant softening length ($r_{\rm soft} = \DeltaP/20$) and the \HACCPM simulation with $\DeltaF=\DeltaP$ for comparison. 
As previously discussed, the \HACCTPM simulation displays significant offsets of the large-scale CDM and baryon power at higher redshift (i.e., $z=10.8$) and extra power on small scales at lower redshift (i.e., $z=2$ and $z=0$)  compared to the reference simulation, whereas the \HACCPM simulation does not show any offsets on large scales but has suppressed power on small scales because of the coarse force resolution. 
The \SFSSOFTEN variation significantly improves the large-scale power spectrum for both species: while CDM no longer shows any notable large-scale offset, the baryon power is still about 1\% suppressed at $z=10.8$, compared to the 12\% suppression in the \HACCTPM run. 
When we completely remove the short-range cross-species interaction in the \SFSOFF option, the offsets on large scales disappear for both CDM and baryons, confirming our initial statement that the large-scale offsets likely arise from the cross-species communication. 
In both the \SFSSOFTEN and \SFSOFF simulations, small-scale structures initially grow at a lower rate than in the \HACCTPM simulation, resulting in less power on large $k$ at higher redshift (e.g., $z=10.8$). This is due to the over-smoothing (\SFSSOFTEN) and the removing (\SFSOFF) of a subset of the short-range forces, essentially reducing the amplitude of the local density perturbation each particle experiences.

The \SFSBOOST variant is added in an effort to compensate the lost power from the missing cross-species interactions by boosting the ``remaining'' short-range forces to the total matter density expected if baryons would perfectly trace CDM. 
Fig.~\ref{fig:convergence_cms} shows that the \SFSBOOST method retains the low-$k$ results of the \SFSOFF variant and does not suffer from discreteness errors. On medium and small scales, the power spectrum is significantly improved. In particular, we can directly compare the total matter power spectrum to the \HACCTPM simulation, for which $P_{m}$ is not affected by the large-scale discreteness errors. We find only small differences at $z=10.8$ and good agreement at $z=2$ and $z=0$. 

We conclude by emphasizing that the \SFSBOOST methodology assumes an idealized approximation that both fluids identically trace the total matter density. This picture is not only inconsistent in multi-fluid simulations, but also is further violated when hydrodynamic forces are additionally included. The \SFSBOOST results are solely intended to highlight the cause of the discreteness errors, where directly removing cross-species interactions can drastically improve the power biases in the simplified gravity-only case we explore here. 
On the other hand, adaptive force softening methods -- akin to our \SFSSOFTEN implementation -- are more generally applicable to multi-species simulations (and hence commonly used in cosmology codes), and are useful mitigation techniques to reduce the erroneous growth, as our results confirm.

\section{Conclusion}
\label{sec:conclusion}

The statistical precision and accuracy of survey observations are increasing steadily, challenging the required predictive capability of cosmological simulations. As the demands on error control become tighter, the long-standing issue of generating accurate multi-species initial conditions needs to be examined closely, especially given the low initial particle sampling density characteristic of cosmological simulations. In this paper, we investigate numerical discreteness errors in gravity-only simulations, where interspecies fluid ICs and evolution can be studied in a simplified setting. Our analysis used measurements of the density power spectrum, a commonly employed diagnostic particularly suited to studies of structure formation irregularities.

We have found that coherently unmixed discretizations of the total density field -- encountered when initializing fluids with simulation tracer particles on offset grids --  lead to biased growth in the individually sampled components if the force resolution employed is finer than the particle spatial sampling scale, i.e., $\DeltaF < \DeltaP$. In single-fluid simulations where the total field is split into species with non-uniform mass, we found that the error is confined to small scales around the (particle sampling) Nyquist mode. 
This behavior can be simply interpreted by examining the asymptotic limit where one species is massless; the suppressed growth of the tracer fluid is understandable given the systematic positional offset from the massive species particles that source the gravitational force. In two-fluid simulations, biased growth occurs out to the largest scales at which the individual transfer functions of the species differ, regardless of whether or not the particle masses of the two species are non-uniform. In both the single and two-fluid cases, the error arises immediately in the simulation and continues growing until the particles become sufficiently mixed due to non-linear gravitational evolution. Subsequent non-linear growth may reduce this error, but significant large-scale biases may still exist at $z = 0$ if proper mitigation strategies are not implemented.

We evaluated several procedures for reducing discreteness errors in two-fluid simulations. First, we examined two previously proposed strategies that involve modifications to the initial condition setup. We found that the large-scale biases in the power spectra for both CDM and baryons are significantly reduced by starting with mixed grid plus glass \citepalias{Bird_2020} or mass-perturbed \citepalias{Hahn_2020} ICs. The pseudo-random glass distribution successfully destroys coherence in the total matter discretization while the mass-perturbed method shifts the discretization to small scales at early times.
Additionally, it is generally advantageous to initialize at later redshifts when the two species are better mixed. We showed that the biased growth on large scales is markedly improved with $z_i \sim 50$ starts for both the standard grid and mass-perturbed ICs. In this case, it is important to use higher order LPT, though multi-species methods incorporating relative velocities beyond first order do not currently exist.

We considered a frozen potential approach as an alternative to starting at lower redshifts while still maintaining offset grid initial conditions. This methodology enforces linear theory evolution up to a specified time, and was also successful in reducing the measured discretization effects. However, avoiding significant suppression of non-linear growth as well as specifying a resolution-dependent redshift parameter complicate its usage.
A final mitigation strategy we examined modified the calculation of short-range interspecies gravitational forces in low-density regions. For example, we found that turning off interspecies interactions within chaining mesh cells below a threshold density effectively removed the large-scale offsets for both CDM and baryons. In general, adaptive softening of gravitational forces reduces the discreteness errors and is a viable approach for addressing this effect.  

The focus of our investigation was entirely devoted to power spectrum analyses, and therefore does not assess the impact of multi-species numerical discreteness errors on other important metrics such as the halo mass function or bispectrum. More work is needed in these areas in order to obtain a deeper understanding of the full impact of this simulation artifact. 
In addition, the simulations presented here were restricted to two species evolved purely under gravity (and, in most cases, treated symmetrically by the force solver). Discreteness errors have the potential to become more pervasive if coupled to a hydrodynamic solver or if a third species such as neutrinos are included. Furthermore, the impact of these effects are not clear when other types of multi-species solvers are considered (e.g., Eulerian or hybrid methodologies). 
In order for cosmological simulations to meet the stringent prediction demands of near-future observational surveys, it is of utmost importance to fully understand and better control the systematics associated with the discrete nature of particle-based simulations.  

\section*{Acknowledgments}

We thank Oliver Hahn, Katrin Heitmann, and Patricia Larsen for helpful discussions and comments on the draft. SH additionally thanks Adrian Melott, Robert Ryne, Sergei Shandarin, and Martin White for past discussions on errors due to discreteness effects. Argonne National Laboratory's work was supported under the U.S. Department of Energy contract DE-AC02-06CH11357. This research was supported by the Exascale Computing Project (17-SC-20-SC), a collaborative effort of the U.S. Department of Energy Office of Science and the National Nuclear Security Administration. This research used resources of the National Energy Research Scientific Computing Center (NERSC), a U.S. Department of Energy Office of Science User Facility located at Lawrence Berkeley National Laboratory, operated under Contract No. DE-AC02-05CH11231 using NERSC award cusp in 2021/22.

\section*{Data Availability}

The analysis data from the simulations presented in this article will be made available upon request. 


\bibliography{references}

\newcommand{\noop}[1]{}
\begin{thebibliography}{}
\makeatletter
\relax
\def\mn@urlcharsother{\let\do\@makeother \do\$\do\&\do\#\do\^\do\_\do\%\do\~}
\def\mn@doi{\begingroup\mn@urlcharsother \@ifnextchar [ {\mn@doi@}
  {\mn@doi@[]}}
\def\mn@doi@[#1]#2{\def\@tempa{#1}\ifx\@tempa\@empty \href
  {http://dx.doi.org/#2} {doi:#2}\else \href {http://dx.doi.org/#2} {#1}\fi
  \endgroup}
\def\mn@eprint#1#2{\mn@eprint@#1:#2::\@nil}
\def\mn@eprint@arXiv#1{\href {http://arxiv.org/abs/#1} {{\tt arXiv:#1}}}
\def\mn@eprint@dblp#1{\href {http://dblp.uni-trier.de/rec/bibtex/#1.xml}
  {dblp:#1}}
\def\mn@eprint@#1:#2:#3:#4\@nil{\def\@tempa {#1}\def\@tempb {#2}\def\@tempc
  {#3}\ifx \@tempc \@empty \let \@tempc \@tempb \let \@tempb \@tempa \fi \ifx
  \@tempb \@empty \def\@tempb {arXiv}\fi \@ifundefined
  {mn@eprint@\@tempb}{\@tempb:\@tempc}{\expandafter \expandafter \csname
  mn@eprint@\@tempb\endcsname \expandafter{\@tempc}}}

\bibitem[\protect\citeauthoryear{{Angulo}, {Hahn}  \& {Abel}}{{Angulo}
  et~al.}{2013}]{angulo2013}
{Angulo} R.~E.,  {Hahn} O.,   {Abel} T.,  2013, \mn@doi [\mnras]
  {10.1093/mnras/stt1135}, \href
  {https://ui.adsabs.harvard.edu/abs/2013MNRAS.434.1756A} {434, 1756}

\bibitem[\protect\citeauthoryear{Bagla \& Padmanabhan}{Bagla \&
  Padmanabhan}{1994}]{Bagla_1994}
Bagla J.~S.,  Padmanabhan T.,  1994, \mn@doi [\mnras]
  {10.1093/mnras/266.1.227}, 266, 227

\bibitem[\protect\citeauthoryear{{Bagla} \& {Prasad}}{{Bagla} \&
  {Prasad}}{2009}]{bagla/prasad:2009}
{Bagla} J.~S.,  {Prasad} J.,  2009, \mn@doi [\mnras]
  {10.1111/j.1365-2966.2008.14224.x}, \href
  {https://ui.adsabs.harvard.edu/abs/2009MNRAS.393..607B} {393, 607}

\bibitem[\protect\citeauthoryear{{Barnes} \& {Hut}}{{Barnes} \&
  {Hut}}{1986}]{barnes/hut:1986}
{Barnes} J.,  {Hut} P.,  1986, \mn@doi [\nat] {10.1038/324446a0}, \href
  {https://ui.adsabs.harvard.edu/abs/1986Natur.324..446B} {324, 446}

\bibitem[\protect\citeauthoryear{{Bate} \& {Burkert}}{{Bate} \&
  {Burkert}}{1997}]{bate/burkert:1997}
{Bate} M.~R.,  {Burkert} A.,  1997, \mn@doi [\mnras]
  {10.1093/mnras/288.4.1060}, \href
  {https://ui.adsabs.harvard.edu/abs/1997MNRAS.288.1060B} {288, 1060}

\bibitem[\protect\citeauthoryear{{Binney} \& {Knebe}}{{Binney} \&
  {Knebe}}{2002}]{binney/knebe:2002}
{Binney} J.,  {Knebe} A.,  2002, \mn@doi [\mnras]
  {10.1046/j.1365-8711.2002.05400.x}, \href
  {https://ui.adsabs.harvard.edu/abs/2002MNRAS.333..378B} {333, 378}

\bibitem[\protect\citeauthoryear{{Bird}, {Feng}, {Pedersen}  \&
  {Font-Ribera}}{{Bird} et~al.}{2020}]{Bird_2020}
{Bird} S.,  {Feng} Y.,  {Pedersen} C.,   {Font-Ribera} A.,  2020, \mn@doi
  [\jcap] {10.1088/1475-7516/2020/06/002}, \href
  {https://ui.adsabs.harvard.edu/abs/2020JCAP...06..002B} {2020, 002}

\bibitem[\protect\citeauthoryear{{Brainerd}, {Scherrer}  \&
  {Villumsen}}{{Brainerd} et~al.}{1993}]{brainerd:1993}
{Brainerd} T.~G.,  {Scherrer} R.~J.,   {Villumsen} J.~V.,  1993, \mn@doi [\apj]
  {10.1086/173417}, \href
  {https://ui.adsabs.harvard.edu/abs/1993ApJ...418..570B} {418, 570}

\bibitem[\protect\citeauthoryear{{DESI Collaboration} et~al.,}{{DESI
  Collaboration} et~al.}{2016}]{desi:2016}
{DESI Collaboration} et~al., 2016, arXiv e-prints, \href
  {https://ui.adsabs.harvard.edu/abs/2016arXiv161100036D} {p. arXiv:1611.00036}

\bibitem[\protect\citeauthoryear{{Diemand}, {Moore}, {Stadel}  \&
  {Kazantzidis}}{{Diemand} et~al.}{2004}]{diemand/etal:2004}
{Diemand} J.,  {Moore} B.,  {Stadel} J.,   {Kazantzidis} S.,  2004, \mn@doi
  [\mnras] {10.1111/j.1365-2966.2004.07424.x}, \href
  {https://ui.adsabs.harvard.edu/abs/2004MNRAS.348..977D} {348, 977}

\bibitem[\protect\citeauthoryear{{Dor{\'e}} et~al.,}{{Dor{\'e}}
  et~al.}{2014}]{spherex:2014}
{Dor{\'e}} O.,  et~al., 2014, arXiv e-prints, \href
  {https://ui.adsabs.harvard.edu/abs/2014arXiv1412.4872D} {p. arXiv:1412.4872}

\bibitem[\protect\citeauthoryear{{Efstathiou} \& {Eastwood}}{{Efstathiou} \&
  {Eastwood}}{1981}]{efstathiou/eastwood:1981}
{Efstathiou} G.,  {Eastwood} J.~W.,  1981, \mn@doi [\mnras]
  {10.1093/mnras/194.3.503}, \href
  {https://ui.adsabs.harvard.edu/abs/1981MNRAS.194..503E} {194, 503}

\bibitem[\protect\citeauthoryear{{Efstathiou}, {Davis}, {White}  \&
  {Frenk}}{{Efstathiou} et~al.}{1985}]{efstathiou/etal:1985}
{Efstathiou} G.,  {Davis} M.,  {White} S.~D.~M.,   {Frenk} C.~S.,  1985,
  \mn@doi [\apjs] {10.1086/191003}, \href
  {https://ui.adsabs.harvard.edu/abs/1985ApJS...57..241E} {57, 241}

\bibitem[\protect\citeauthoryear{{Emberson}, {Frontiere}, {Habib}, {Heitmann},
  {Larsen}, {Finkel}  \& {Pope}}{{Emberson} et~al.}{2019}]{JD2019}
{Emberson} J.~D.,  {Frontiere} N.,  {Habib} S.,  {Heitmann} K.,  {Larsen} P.,
  {Finkel} H.,   {Pope} A.,  2019, \mn@doi [\apj] {10.3847/1538-4357/ab1b31},
  \href {https://ui.adsabs.harvard.edu/abs/2019ApJ...877...85E} {877, 85}

\bibitem[\protect\citeauthoryear{Feng, Bird, Anderson, Font-Ribera  \&
  Pedersen}{Feng et~al.}{2018}]{yu_feng_2018_1451799}
Feng Y.,  Bird S.,  Anderson L.,  Font-Ribera A.,   Pedersen C.,  2018,
  MP-Gadget/MP-Gadget: A tag for getting a DOI,
  \mn@doi{10.5281/zenodo.1451799}, \url
  {https://doi.org/10.5281/zenodo.1451799}

\bibitem[\protect\citeauthoryear{Gafton \& Rosswog}{Gafton \&
  Rosswog}{2011}]{gafton&rosswog:2011}
Gafton E.,  Rosswog S.,  2011, \mn@doi [\mnras]
  {https://doi.org/10.1111/j.1365-2966.2011.19528.x}, 418, 770

\bibitem[\protect\citeauthoryear{{G{\"o}tz} \& {Sommer-Larsen}}{{G{\"o}tz} \&
  {Sommer-Larsen}}{2003}]{gotz/sommerlarsen:2003}
{G{\"o}tz} M.,  {Sommer-Larsen} J.,  2003, \mn@doi [\apss]
  {10.1023/A:1024073909753}, \href
  {https://ui.adsabs.harvard.edu/abs/2003Ap&SS.284..341G} {284, 341}

\bibitem[\protect\citeauthoryear{{Habib} et~al.,}{{Habib}
  et~al.}{2016}]{Habib2016}
{Habib} S.,  et~al., 2016, \mn@doi [\na] {10.1016/j.newast.2015.06.003}, \href
  {https://ui.adsabs.harvard.edu/abs/2016NewA...42...49H} {42, 49}

\bibitem[\protect\citeauthoryear{Hahn \& Angulo}{Hahn \&
  Angulo}{2015}]{Hahn_2015}
Hahn O.,  Angulo R.~E.,  2015, \mn@doi [\mnras] {10.1093/mnras/stv2304}, 455,
  1115

\bibitem[\protect\citeauthoryear{Hahn, Rampf  \& Uhlemann}{Hahn
  et~al.}{2020}]{Hahn_2020}
Hahn O.,  Rampf C.,   Uhlemann C.,  2020, \mn@doi [\mnras]
  {10.1093/mnras/staa3773}, 503, 426

\bibitem[\protect\citeauthoryear{{Hamana}, {Yoshida}  \& {Suto}}{{Hamana}
  et~al.}{2002}]{hamana/etal:2002}
{Hamana} T.,  {Yoshida} N.,   {Suto} Y.,  2002, \mn@doi [\apj]
  {10.1086/338970}, \href
  {https://ui.adsabs.harvard.edu/abs/2002ApJ...568..455H} {568, 455}

\bibitem[\protect\citeauthoryear{{Heitmann}, {Ricker}, {Warren}  \&
  {Habib}}{{Heitmann} et~al.}{2005}]{heitmann2005ApJS}
{Heitmann} K.,  {Ricker} P.~M.,  {Warren} M.~S.,   {Habib} S.,  2005, \mn@doi
  [\apjs] {10.1086/432646}, \href
  {https://ui.adsabs.harvard.edu/abs/2005ApJS..160...28H} {160, 28}

\bibitem[\protect\citeauthoryear{{Heitmann}, {White}, {Wagner}, {Habib}  \&
  {Higdon}}{{Heitmann} et~al.}{2010}]{heitmann2010ApJ}
{Heitmann} K.,  {White} M.,  {Wagner} C.,  {Habib} S.,   {Higdon} D.,  2010,
  \mn@doi [\apj] {10.1088/0004-637X/715/1/104}, \href
  {https://ui.adsabs.harvard.edu/abs/2010ApJ...715..104H} {715, 104}

\bibitem[\protect\citeauthoryear{{Hockney} \& {Eastwood}}{{Hockney} \&
  {Eastwood}}{1988}]{1988book}
{Hockney} R.~W.,  {Eastwood} J.~W.,  1988, {Computer simulation using
  particles}.
{CRC Press}

\bibitem[\protect\citeauthoryear{{Hopkins} et~al.,}{{Hopkins}
  et~al.}{2018}]{hopkins/etal:2018}
{Hopkins} P.~F.,  et~al., 2018, \mn@doi [\mnras] {10.1093/mnras/sty1690}, \href
  {https://ui.adsabs.harvard.edu/abs/2018MNRAS.480..800H} {480, 800}

\bibitem[\protect\citeauthoryear{{Ivezi{\'c}} et~al.,}{{Ivezi{\'c}}
  et~al.}{2019}]{lsst:2019}
{Ivezi{\'c}} {\v{Z}}.,  et~al., 2019, \mn@doi [\apj]
  {10.3847/1538-4357/ab042c}, \href
  {https://ui.adsabs.harvard.edu/abs/2019ApJ...873..111I} {873, 111}

\bibitem[\protect\citeauthoryear{{Joyce}, {Marcos}  \& {Baertschiger}}{{Joyce}
  et~al.}{2009}]{joyce/etal:2009}
{Joyce} M.,  {Marcos} B.,   {Baertschiger} T.,  2009, \mn@doi [\mnras]
  {10.1111/j.1365-2966.2008.14290.x}, \href
  {https://ui.adsabs.harvard.edu/abs/2009MNRAS.394..751J} {394, 751}

\bibitem[\protect\citeauthoryear{{Klypin} \& {Shandarin}}{{Klypin} \&
  {Shandarin}}{1983}]{klypin1983}
{Klypin} A.~A.,  {Shandarin} S.~F.,  1983, \mn@doi [\mnras]
  {10.1093/mnras/204.3.891}, \href
  {https://ui.adsabs.harvard.edu/abs/1983MNRAS.204..891K} {204, 891}

\bibitem[\protect\citeauthoryear{{Knebe}, {Kravtsov}, {Gottl{\"o}ber}  \&
  {Klypin}}{{Knebe} et~al.}{2000}]{knebe/etal:2000}
{Knebe} A.,  {Kravtsov} A.~V.,  {Gottl{\"o}ber} S.,   {Klypin} A.~A.,  2000,
  \mn@doi [\mnras] {10.1046/j.1365-8711.2000.03673.x}, \href
  {https://ui.adsabs.harvard.edu/abs/2000MNRAS.317..630K} {317, 630}

\bibitem[\protect\citeauthoryear{{Laureijs} et~al.,}{{Laureijs}
  et~al.}{2011}]{euclid:2011}
{Laureijs} R.,  et~al., 2011, arXiv e-prints, \href
  {https://ui.adsabs.harvard.edu/abs/2011arXiv1110.3193L} {p. arXiv:1110.3193}

\bibitem[\protect\citeauthoryear{{Lewis}, {Challinor}  \& {Lasenby}}{{Lewis}
  et~al.}{2000}]{Lewis2000}
{Lewis} A.,  {Challinor} A.,   {Lasenby} A.,  2000, \mn@doi [\apj]
  {10.1086/309179}, \href
  {https://ui.adsabs.harvard.edu/abs/2000ApJ...538..473L} {538, 473}

\bibitem[\protect\citeauthoryear{{Little}, {Weinberg}  \& {Park}}{{Little}
  et~al.}{1991}]{little/etal:1991}
{Little} B.,  {Weinberg} D.~H.,   {Park} C.,  1991, \mn@doi [\mnras]
  {10.1093/mnras/253.2.295}, \href
  {https://ui.adsabs.harvard.edu/abs/1991MNRAS.253..295L} {253, 295}

\bibitem[\protect\citeauthoryear{{Ludlow}, {Schaye}, {Schaller}  \&
  {Richings}}{{Ludlow} et~al.}{2019}]{ludlow/etal:2019}
{Ludlow} A.~D.,  {Schaye} J.,  {Schaller} M.,   {Richings} J.,  2019, \mn@doi
  [\mnras] {10.1093/mnrasl/slz110}, \href
  {https://ui.adsabs.harvard.edu/abs/2019MNRAS.488L.123L} {488, L123}

\bibitem[\protect\citeauthoryear{{McCarthy}, {Schaye}, {Bird}  \& {Le
  Brun}}{{McCarthy} et~al.}{2017}]{bahamas:2017}
{McCarthy} I.~G.,  {Schaye} J.,  {Bird} S.,   {Le Brun} A. M.~C.,  2017,
  \mn@doi [\mnras] {10.1093/mnras/stw2792}, \href
  {https://ui.adsabs.harvard.edu/abs/2017MNRAS.465.2936M} {465, 2936}

\bibitem[\protect\citeauthoryear{{Melott}}{{Melott}}{2007}]{melott:2007}
{Melott} A.~L.,  2007, arXiv e-prints, \href
  {https://ui.adsabs.harvard.edu/abs/2007arXiv0709.0745M} {p. arXiv:0709.0745}

\bibitem[\protect\citeauthoryear{{Melott}, {Shandarin}, {Splinter}  \&
  {Suto}}{{Melott} et~al.}{1997}]{melott/etal:1997}
{Melott} A.~L.,  {Shandarin} S.~F.,  {Splinter} R.~J.,   {Suto} Y.,  1997,
  \mn@doi [\apjl] {10.1086/310590}, \href
  {https://ui.adsabs.harvard.edu/abs/1997ApJ...479L..79M} {479, L79}

\bibitem[\protect\citeauthoryear{Michaux, Hahn, Rampf  \& Angulo}{Michaux
  et~al.}{2020}]{Michaux_2020}
Michaux M.,  Hahn O.,  Rampf C.,   Angulo R.~E.,  2020, \mn@doi [\mnras]
  {10.1093/mnras/staa3149}, 500, 663

\bibitem[\protect\citeauthoryear{Nicholson}{Nicholson}{1983}]{nicholson}
Nicholson D.,  1983, Introduction to Plasma Theory.
Wiley, \url {https://books.google.com/books?id=fyRRAAAAMAAJ}

\bibitem[\protect\citeauthoryear{{O'Leary} \& {McQuinn}}{{O'Leary} \&
  {McQuinn}}{2012}]{oleary/mcquinn:2012}
{O'Leary} R.~M.,  {McQuinn} M.,  2012, \mn@doi [\apj]
  {10.1088/0004-637X/760/1/4}, \href
  {https://ui.adsabs.harvard.edu/abs/2012ApJ...760....4O} {760, 4}

\bibitem[\protect\citeauthoryear{{Okuda} \& {Birdsall}}{{Okuda} \&
  {Birdsall}}{1970}]{okuda1970}
{Okuda} H.,  {Birdsall} C.~K.,  1970, \mn@doi [Physics of Fluids]
  {10.1063/1.1693210}, \href
  {https://ui.adsabs.harvard.edu/abs/1970PhFl...13.2123O} {13, 2123}

\bibitem[\protect\citeauthoryear{{Peebles}}{{Peebles}}{2021}]{peebles:1993}
{Peebles} P.~J.~E.,  2021, {Principles of Physical Cosmology}.
{Princeton University Press}, \mn@doi{10.1515/9780691206721}

\bibitem[\protect\citeauthoryear{{Planck Collaboration}}{{Planck
  Collaboration}}{2020}]{Planck2020}
{Planck Collaboration} 2020, \mn@doi [\aap] {10.1051/0004-6361/201833910},
  \href {https://ui.adsabs.harvard.edu/abs/2020A&A...641A...6P} {641, A6}

\bibitem[\protect\citeauthoryear{{Power}, {Navarro}, {Jenkins}, {Frenk},
  {White}, {Springel}, {Stadel}  \& {Quinn}}{{Power}
  et~al.}{2003}]{power/etal:2003}
{Power} C.,  {Navarro} J.~F.,  {Jenkins} A.,  {Frenk} C.~S.,  {White} S.~D.~M.,
   {Springel} V.,  {Stadel} J.,   {Quinn} T.,  2003, \mn@doi [\mnras]
  {10.1046/j.1365-8711.2003.05925.x}, \href
  {https://ui.adsabs.harvard.edu/abs/2003MNRAS.338...14P} {338, 14}

\bibitem[\protect\citeauthoryear{{Power}, {Robotham}, {Obreschkow}, {Hobbs}  \&
  {Lewis}}{{Power} et~al.}{2016}]{power/etal:2016}
{Power} C.,  {Robotham} A.~S.~G.,  {Obreschkow} D.,  {Hobbs} A.,   {Lewis}
  G.~F.,  2016, \mn@doi [\mnras] {10.1093/mnras/stw1644}, \href
  {https://ui.adsabs.harvard.edu/abs/2016MNRAS.462..474P} {462, 474}

\bibitem[\protect\citeauthoryear{{Romeo}, {Agertz}, {Moore}  \&
  {Stadel}}{{Romeo} et~al.}{2008}]{romeo/etal:2008}
{Romeo} A.~B.,  {Agertz} O.,  {Moore} B.,   {Stadel} J.,  2008, \mn@doi [\apj]
  {10.1086/591236}, \href
  {https://ui.adsabs.harvard.edu/abs/2008ApJ...686....1R} {686, 1}

\bibitem[\protect\citeauthoryear{{Schaye} et~al.,}{{Schaye}
  et~al.}{2015}]{eagle:2015}
{Schaye} J.,  et~al., 2015, \mn@doi [\mnras] {10.1093/mnras/stu2058}, \href
  {https://ui.adsabs.harvard.edu/abs/2015MNRAS.446..521S} {446, 521}

\bibitem[\protect\citeauthoryear{Schneider et~al.,}{Schneider
  et~al.}{2016}]{Schneider_2016}
Schneider A.,  et~al., 2016, \mn@doi [\jcap] {10.1088/1475-7516/2016/04/047},
  2016, 047

\bibitem[\protect\citeauthoryear{{Spergel} et~al.,}{{Spergel}
  et~al.}{2015}]{roman:2015}
{Spergel} D.,  et~al., 2015, \mn@doi [arXiv e-prints]
  {10.48550/arXiv.1503.03757}, \href
  {https://ui.adsabs.harvard.edu/abs/2015arXiv150303757S} {p. arXiv:1503.03757}

\bibitem[\protect\citeauthoryear{{Splinter}, {Melott}, {Shandarin}  \&
  {Suto}}{{Splinter} et~al.}{1998}]{splinter/etal:1998}
{Splinter} R.~J.,  {Melott} A.~L.,  {Shandarin} S.~F.,   {Suto} Y.,  1998,
  \mn@doi [\apj] {10.1086/305450}, \href
  {https://ui.adsabs.harvard.edu/abs/1998ApJ...497...38S} {497, 38}

\bibitem[\protect\citeauthoryear{{Steinmetz} \& {White}}{{Steinmetz} \&
  {White}}{1997}]{steinmetz/white:1997}
{Steinmetz} M.,  {White} S. D.~M.,  1997, \mn@doi [\mnras]
  {10.1093/mnras/288.3.545}, \href
  {https://ui.adsabs.harvard.edu/abs/1997MNRAS.288..545S} {288, 545}

\bibitem[\protect\citeauthoryear{{Sullivan}, {Emberson}, {Habib}  \&
  {Frontiere}}{{Sullivan} et~al.}{2023}]{sullivan/etal:2023}
{Sullivan} J.~M.,  {Emberson} J.~D.,  {Habib} S.,   {Frontiere} N.,  2023,
  arXiv e-prints, \href {https://ui.adsabs.harvard.edu/abs/2023arXiv230209134S}
  {p. arXiv:2302.09134}

\bibitem[\protect\citeauthoryear{{Valkenburg} \&
  {Villaescusa-Navarro}}{{Valkenburg} \&
  {Villaescusa-Navarro}}{2017}]{valkenburg/etal:2017}
{Valkenburg} W.,  {Villaescusa-Navarro} F.,  2017, \mn@doi [\mnras]
  {10.1093/mnras/stx376}, \href
  {https://ui.adsabs.harvard.edu/abs/2017MNRAS.467.4401V} {467, 4401}

\bibitem[\protect\citeauthoryear{Villaescusa-Navarro, Banerjee, Dalal,
  Castorina, Scoccimarro, Angulo  \& Spergel}{Villaescusa-Navarro
  et~al.}{2018}]{Villaescusa_Navarro_2018}
Villaescusa-Navarro F.,  Banerjee A.,  Dalal N.,  Castorina E.,  Scoccimarro
  R.,  Angulo R.,   Spergel D.~N.,  2018, \mn@doi [\apj]
  {10.3847/1538-4357/aac6bf}, 861, 53

\bibitem[\protect\citeauthoryear{{Vogelsberger} et~al.,}{{Vogelsberger}
  et~al.}{2014}]{illustris:2014}
{Vogelsberger} M.,  et~al., 2014, \mn@doi [\nat] {10.1038/nature13316}, \href
  {https://ui.adsabs.harvard.edu/abs/2014Natur.509..177V} {509, 177}

\bibitem[\protect\citeauthoryear{{Vra{\v{s}}til} \& {Habib}}{{Vra{\v{s}}til} \&
  {Habib}}{2020}]{vrastil:2020}
{Vra{\v{s}}til} M.,  {Habib} S.,  2020, \mn@doi [\mnras]
  {10.1093/mnras/staa333}, \href
  {https://ui.adsabs.harvard.edu/abs/2020MNRAS.493.2085V} {493, 2085}

\bibitem[\protect\citeauthoryear{{Wang} \& {White}}{{Wang} \&
  {White}}{2007}]{wang/white:2007}
{Wang} J.,  {White} S. D.~M.,  2007, \mn@doi [\mnras]
  {10.1111/j.1365-2966.2007.12053.x}, \href
  {https://ui.adsabs.harvard.edu/abs/2007MNRAS.380...93W} {380, 93}

\bibitem[\protect\citeauthoryear{White}{White}{1994}]{glass_note}
White S. D.~M.,  1994, Formation and Evolution of Galaxies: Les Houches
  Lectures, \mn@doi{10.48550/ARXIV.ASTRO-PH/9410043}, \url
  {https://arxiv.org/abs/astro-ph/9410043}

\bibitem[\protect\citeauthoryear{{White}}{{White}}{2014}]{white:2014}
{White} M.,  2014, \mn@doi [\mnras] {10.1093/mnras/stu209}, \href
  {https://ui.adsabs.harvard.edu/abs/2014MNRAS.439.3630W} {439, 3630}

\bibitem[\protect\citeauthoryear{Yoshida, Sugiyama  \& Hernquist}{Yoshida
  et~al.}{2003}]{Yoshida_2003}
Yoshida N.,  Sugiyama N.,   Hernquist L.,  2003, \mn@doi [\mnras]
  {10.1046/j.1365-8711.2003.06829.x}, 344, 481

\bibitem[\protect\citeauthoryear{{Zel'dovich}}{{Zel'dovich}}{1970}]{Zeldovich1970}
{Zel'dovich} Y.~B.,  1970, \aap, \href
  {https://ui.adsabs.harvard.edu/abs/1970A&A.....5...84Z} {5, 84}

\makeatother
\end{thebibliography}
\bibliographystyle{mnras}

\appendix
\section{Two-fluid Plane Wave Experiment}
\label{app:planewave}

We provide here an illustrative example of how an unmixed discretization of the total matter field leads to biased growth in two-fluid simulations. This is accomplished using an idealized setup where two particle types (``CDM'' and ``baryon'') of uniform mass are initialized using a single plane wave density perturbation with a species-dependent amplitude. We find that the species with the larger (smaller) perturbation amplitude exhibits enhanced (suppressed) growth due to paired particle coupling producing an inward (outward) force around peaks in the total matter density field. If we add perturbations to other scales, but with an equal amplitude for both species, then the biased growth remains confined only to the scale with the species-dependent amplitude. In other words, biased growth is a general feature of unmixed two-fluid simulations that manifests on scales where the individual transfer functions differ, consistent with the observations in Section~\ref{sec:unequaltransfer}.

We consider two simulations where CDM and baryon particles of uniform mass are evolved in an EdS universe ($\Omega_c=\Omega_b=0.5$; $h = 1$): 1) a single-mode plane wave and 2) a three-mode perturbation. In both cases, the perturbations are aligned with the $x$-axis. Starting from two offset $128^3$ grids for baryons and CDM in an $L=32\Mpch$ box, we apply the ZA for an initial potential of the form $\phi^\alpha(q_x) = \sum_i A^\alpha_i \cos(k_i q_x)$, with $\alpha \in \{{c}, {b}\}$ denoting CDM or baryons, and where $k_i = 2\pi i/L$ for $i>0$ are the wave vectors.
The amplitudes $A^\alpha_i$ are chosen so that the initial conditions at $z_i=200$ remain in the linear regime, i.e., well before shell-crossing occurs. For the single-mode simulation, we fix $A^\alpha_i=0$ except for $i=5$ and choose $A^{c}_5=6.1 \times 10^4\;  (\mathrm{km}/\mathrm{s})^2$. To imitate the effect of separate transfer functions, the baryon amplitude is set to $A^{b}=A^{c}/4$. For the second setup, we select the $i=(2,5,9)$ modes to be non-zero. We set $A^{c}$ to $(9.5,15,8.5) \times 10^3\; (\mathrm{km}/\mathrm{s})^2$ for the CDM perturbations and $A^{b}=(A^{c}_2, A^{c}_5/4, A^{c}_9)$ for the baryons, i.e., only the $i=5$ baryon mode has a different amplitude. We then compute the gravitational accelerations on each particle for $\DeltaF=\DeltaP$ and $\DeltaF=\DeltaP/4$ using the \HACCPM force solver in the first time step.

\begin{figure*}
\includegraphics[width=\linewidth]{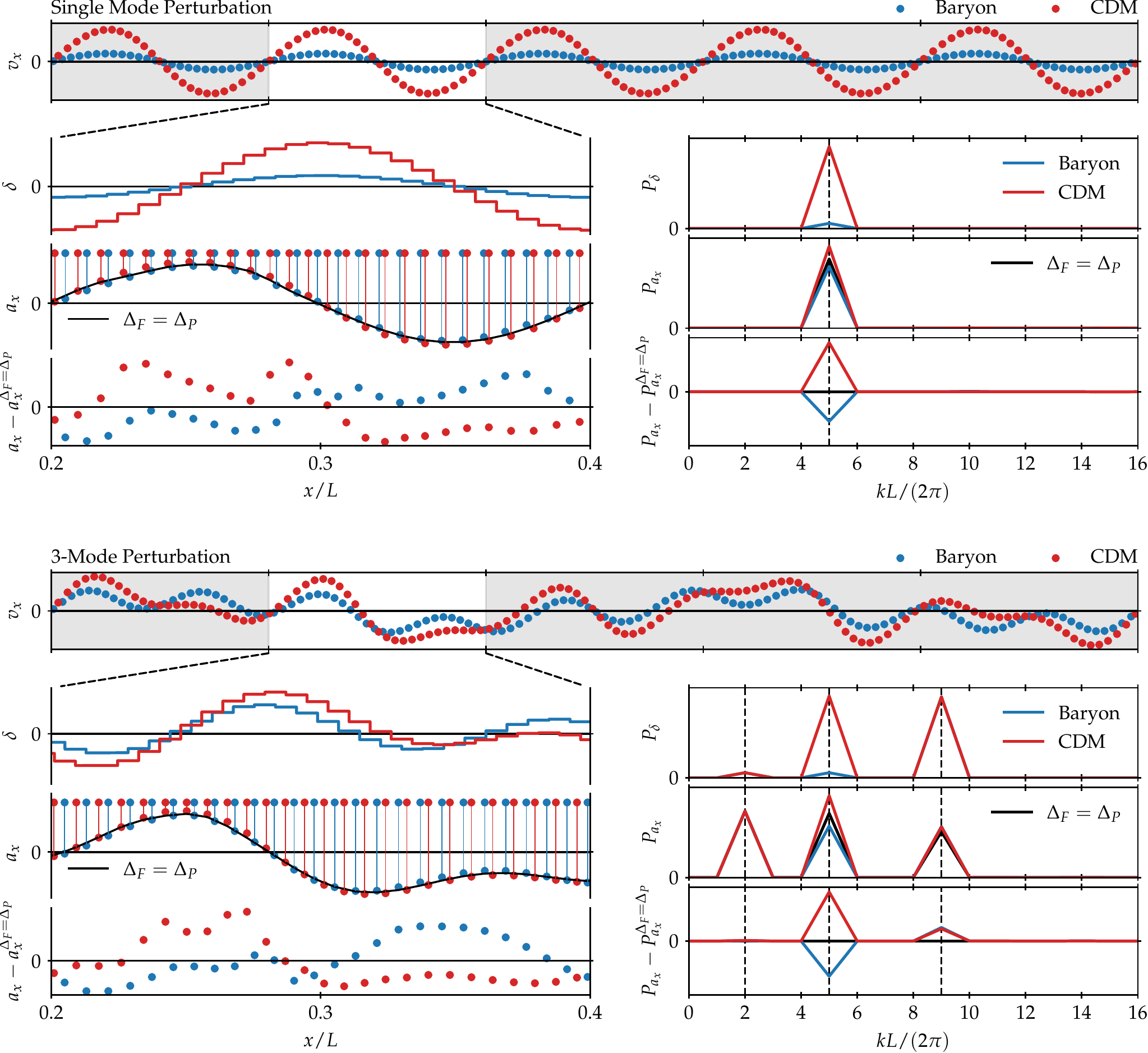}
\caption{Comparison of the particle acceleration between different force resolutions in two idealized simulation setups: a single-mode plane wave (top) and a three-mode perturbation (bottom). In both cases, the $k=5\times 2\pi/L$ mode for the baryon species is set to a lower amplitude than the CDM perturbation. From top to bottom, each subfigure shows the phase-space distribution, the density distribution measured on a grid with $\DeltaG=\DeltaP$, the acceleration measured at the particle position, and the acceleration difference between the \HACCPM force solver with $\DeltaF=\DeltaP/4$ and $\DeltaF=\DeltaP$. The density and acceleration panels are shown for a subvolume of the simulation box (left) and in Fourier space (right). In order to highlight the relative positions between CDM and baryon particles, we line up the particles next to each other above the accelerations.}
\label{fig:plane_waves}
\end{figure*}

The particle configurations and acceleration measurements are illustrated in Fig.~\ref{fig:plane_waves}. Starting with the plane wave setup, we find that the particle accelerations for baryons are damped (i.e., closer to zero) on both sides of the density peak for the $\DeltaF=\DeltaP/4$ force resolution, whereas the CDM accelerations are boosted. This also manifests itself in Fourier space, where we find a decrease (increase) in the amplitude of the acceleration spectrum for baryons (CDM) at $k=5\times 2\pi/L$ when increasing the force resolution. Notably, for $\DeltaF=\DeltaP$, the CDM and baryon acceleration spectrum is equivalent, as is expected if both species describe continuous fluids. Discreteness errors appear when sampling the forces on a finer resolution, and in the case of this toy model, the opposite effect on CDM and baryons can be understood when looking at the relative position of CDM and baryon particles around the density peaks. Due to the different displacement amplitudes for the individual species, the CDM and baryon particles form ``pairs'', where the CDM particle is always on the outside with respect to the density peak. If the force resolution is fine enough to resolve the 2-body interactions, this alignment will enhance (suppress) the inward acceleration for CDM (baryons), boosting (damping) the growth of that specific mode. This consistent arrangement occurs as long as the displacements are small, i.e., at high redshift. For larger perturbations, the particle ordering will eventually invert at some positions.

We show the same statistics for the three-mode configuration in the lower panel of Fig \ref{fig:plane_waves}. We again find that the acceleration amplitudes for the individual species are coherently biased on the scale of the $k=5\times 2\pi/L$ mode when using the finer force resolution. This is the only mode that was initialized with different amplitudes for CDM and baryons. In the acceleration spectrum we also notice that the $k=9\times 2\pi/L$ mode is higher for both species. Since this is a species-independent feature, it is likely of a different origin than the asymmetric biased growth seen at $k=5\times 2\pi/L$. This was confirmed using a separate simulation where we used equal baryon and CDM amplitudes for all three modes, in which case we found the same difference in $P_{a_x}$ between force resolutions. Hence, the ``bump'' is merely caused by under-resolving this mode fully with $\DeltaF=\DeltaP$, and not a manifestation of discreteness errors in two-fluid simulations.

These toy models also provide us with some insight into how the different mitigation strategies (discussed in Section~\ref{sec:convergence}) circumvent the large-scale discreteness errors. Glass configurations such as used in \citetalias{Bird_2020} add noise to the pre-IC positions which destroys the consistent arrangement of baryon and CDM particle pairs. Using the total matter transfer function to calculate the particle displacements \citepalias{Hahn_2020} similarly prevents the formation of pairs by offsetting the two species coherently. If the simulation is initialized at lower redshift, the displacement amplitudes compared to the pre-IC interparticle separation is larger, allowing some CDM particles to ``overtake'' baryon particles and preventing a coherent acceleration bias on the scale of the $k$-mode where the initial amplitudes are different. Finally, softening or removing the short-range interspecies interactions removes the biased accelerations originating from two-body interactions.
 
We close this section by emphasizing that the two configurations presented are highly idealized, with specially selected modes and amplitudes used solely for illustrative purposes. Cosmological simulations are in general more complicated with biases introduced on all relevant modes, where interpretation of error growth is more difficult.
That being said, the simple case-study does provide a consistent picture with the cosmological two-fluid simulation results, where discretization errors are confined to the scales at which the individual transfer functions are different.

\label{lastpage}
\end{document}